\documentclass[preprint]{revtex4}
\usepackage[dvipdfmx]{graphicx}
\usepackage{ulem}
\begin{document}


\title{Equilibrium nuclear ensembles taking into account vaporization of hot nuclei in dense stellar matter }


\author{Shun Furusawa}
\email{furusawa@fias.uni-frankfurt.de}
\affiliation{Interdisciplinary Theoretical Science (iTHES) Research Group, RIKEN, Wako, Saitama 351-0198, Japan}
\affiliation{Frankfurt Institute for Advanced Studies, J.W. Goethe University, 60438 Frankfurt am Main, Germany}
\author{Igor Mishustin}
\affiliation{Frankfurt Institute for Advanced Studies, J.W. Goethe University, 60438 Frankfurt am Main, Germany}
\affiliation{Russian  Research Center Kurchatov Institute, Moscow 123182, Russia
}


\date{\today}

\begin{abstract}
We investigate the high-temperature effect on the nuclear matter that
 consists of mixture of  nucleons and all nuclei in the dense and hot stellar environment.  
The individual nuclei are described within the compressible liquid-drop model that is 
based on Skyrme interactions for bulk energies and that takes into account modifications of the surface and Coulomb energies at finite temperatures and densities. 
The free-energy density is minimized with respect to  the individual equilibrium
 densities of all heavy nuclei and  the nuclear composition. We find that their optimized equilibrium densities become smaller and smaller
 at high temperatures because of the increase
 of thermal contributions to bulk free energies and  the reduction of  surface energies. 
The neutron-rich nuclei become unstable and disappear one after another at some temperatures.   
The calculations are performed for two sets of model parameters 
leading  to different values of the slope parameter in the nuclear symmetry energy.
It is found that the larger slope parameter 
 reduces the equilibrium densities and the melting temperatures.
We also compare the new model with some other approaches and find that the mass fractions of heavy nuclei in the previous calculations that omit vaporization are underestimated at $T \lesssim 10$~MeV and overestimated at $T\gtrsim10$~MeV. 
The further sophistication of calculations of nuclear vaporization and of light clusters would be required to construct the equation of state for  explosive astrophysical phenomena.
\end{abstract}


\maketitle

\section{Introduction \label{intro}}
Hot and dense matter can be realized  both in terrestrial experiments of heavy ion collision and explosive astrophysical phenomena such as core collapse supernovae and mergers of compact stars  (see e.g., Refs  \cite{botvina04,janka12,kotake12,burrows13,foglizzo15,shibata11,faber12}). 
Even after various endeavors of the previous decades, we have little information about  the  nuclear equations of state (EOSs)  both at sub and supra nuclear densities
 and, up to now, there is no consensus about its theoretical description
in the nuclear community \cite{oertel17}.
This is partially because 
it is difficult  to  reproduce
the density, temperature and proton-neutron ratio dependences of  nuclear matter arbitrarily in terrestrial laboratories, although several experimental efforts are underway. 
Furthermore, it cannot be calculated from first-principles due to enormous  complexity.

The EOS provides information on  nuclear composition 
 in addition to thermodynamical quantities.
The former plays an important role to alter the charge fraction of stellar matter  
 through weak interactions \cite{raduta16, raduta17, furusawa17b}
and, as a result,  affects the dynamics and  synthesis of heavy elements
in  these events  \cite{hix03,lentz12,wanajo14, sekiguchi15}.
The latter is directly linked to the structure of compact stars and the dynamics of 
their formation and merger.

In the literature, 
the single nucleus approximation (SNA)  or the multi nucleus approximation (MNA) 
is used to construct huge EOS data covering a wide range of density, temperature and proton fraction for stellar matter \cite{buyukcizmeci13,sheng11}.
In the  former,  the ensemble of heavy nuclei is represented by a single nucleus
\cite{lattimer91, shen98a,shen98b,shen11,togashi17},
which is calculated by the compressible liquid drop model  or Thomas Fermi model. 
In such calculations, 
 the optimizations for nuclear structure, such as compression, are taken into consideration, whereas the approach is not able to  provide a realistic nuclear composition.
Furthermore,  the mass number and mass fraction of the representative nucleus 
are considerably deviated from the actual values obtained for full nuclear ensemble  \cite{furusawa17c}. 
{
In the latter type of  EOSs such as  SMSM \cite{botvina04, botvina10,buyukcizmeci14}, HS  \cite{hempel10, steiner13} and FYSS \cite{furusawa11,furusawa13a, furusawa17a,furusawa17d} EOSs,  the nuclear composition including all nuclear species is optimized for each set of thermodynamical condition.}
On the other hand, the MNA EOSs 
use very rough approximations, such as an incompressible-liquid drop model 
for nuclear binding energies  at high densities and temperatures,  to reduce the numerical cost for calculations of the full statistical ensemble. 
The fully self-consistent calculation with the multi-nucleus composition 
including in-medium effects on nuclear structure can hardly be performed due to the complexity \cite{gulminelli15,furusawa17c}. 

The purpose of this series of papers is to investigate what happens in the full nuclear ensemble as
 the density or temperature increases.
In our previous paper \cite{furusawa17c}, we took into account self-consistently the compression or decompression of heavy nuclei in the multi-nucleus description, i.e., the changes of equilibrium densities of individual nuclei embedded in a dense stellar environment. We then found that heavy nuclei in the ensemble are either decompressed or
compressed depending on whether nucleons are dripped. Furthermore, we demonstrated that the EOS with MNA shows better agreement with the self-consistent calculation of the mass fraction and average mass number of heavy nuclei compared with the EOS with SNA below about  $0.3$ times the nuclear saturation density $n_0$
and low temperatures. 

In this paper, we investigate the self-consistent multi-nucleus system at higher temperatures up to $T \sim$15~MeV. 
In the previous paper, we discussed  the results below $T= 3$ MeV, because
the bulk energy was estimated by the simple parametric formula of the bulk energies of heavy nuclei, which is valid only at low temperatures and around the saturation density.
In this work, we replace it by a more reliable calculation based on a Skyrme-type mean-field interaction (see Refs. \cite{khodel87,smirnov88,oyamatsu93,fayans00,oyamatsu03})
and the exact expression for thermal excitation energies. 
The improvement allows us to discuss the vaporization of  nuclear species at the temperatures, 
at which the equilibrium densities can not be obtained and the nuclei disappear one after another.
{
Such self-consistent calculations of nuclear abundance and dissolution have not been reported so far in the framework of the multi-nucleus description, whereas nuclear structures  at high temperatures are investigated mainly in SNA (see, e.g., \cite{newton09}).
Gulminelli and Oertel \cite{gulminelli15} investigated 
the equilibrium between nucleons and nuclei as well as nuclear composition in the multi nucleus description, although the optimization of nuclear equilibrium density is not taken into consideration.}

This work is organized as follows:
In Sec.~\ref{sec:model}, we formulate a model of EOS,
which is used  to find the nuclear abundance and to take into account the possibility  of nuclear vaporization. 
The results at  some typical astrophysical conditions for two different parameter sets for bulk nuclear matter
 are discussed in Sec.~\ref{sec:res1}.
{
In addition, we provide systematical comparisons with other models for free energies of heavy and light nuclei in Sec.~\ref{sec:res2}.}
Section~\ref{sec:conc} is devoted to the conclusion. 

\section{Model \label{sec:model}}
The basis of our EOS  model is 
the same as the previous one. For details, we refer the readers to Furusawa and Mishustin \cite{furusawa17c}.
The  EOS as a function of baryon density $n_B$,  temperature $T$, and total proton fraction $Y_p$  
is obtained by optimizing the nuclear composition as well as the nuclear equilibration, or minimizing the free energy density with respect to the number densities  of all particles and equilibrium densities of heavy nuclei.
The free energy density 
 is expressed as
\begin{eqnarray}
f = f_p + f_n  +  \sum_j  n_j (F^t_{j} + M_{j})+ \sum_i  n_i (F^t_{i} + M_{i}) ,
\label{total}
\end{eqnarray}
where $f_p$ and $f_n$  are the free energy densities of free protons and neutrons  and $n_{i/j}$,
$F^t_{i/j}$ and $M_{i/j}$ are the number densities, 
translational free energies and mass free energies of  light clusters $j$  and heavy nuclei $i$.
{
For light clusters with  the atomic number  $Z_j \leq5$ or  the neutron number $N_j \leq5$, we include only the nuclide with the  available experimental mass data, 
while  all heavy nuclei with  $6 \leq Z_i \leq 2000$ and  $6 \leq N_i \leq 2000$ are taken into consideration as long as they are stable. The maximum asymmetry of heavy nuclei, $\delta=|1-2Z_i/A_i|$, is around 0.8, which  corresponds to extremely neutron-rich nuclei such as $(N_i,Z_i)=(200, 1800)$.} 
{
The nuclear interaction of bulk nuclear matter is explicitly taken into account in  $f_{p/n}$ and $M_i$.  
A compressible liquid drop model is adopted for $M_i$, which is described by the sum of bulk, Coulomb and surface free energies as $M_i=F_i^B+F_i^C+F_i^S$.
For $M_j$, experimental mass data with Coulomb shifts are utilized. 
The interactions among different species are represented by the excluded volume effects in $f_{p/n}$ and $F^t_{i/j}$.}
Below we explain each term of the model free energy.

We improve our previous model by replacing the parametric expression of the nuclear bulk energies by  the more realistic calculation based on Skyrme type interactions, which  
 provides the free energy per baryon of uniform nuclear matter, $\omega$, as a function of local values of  $n_B$, charge fraction $x$, and $T$. The charge fraction means  the  ratio of total proton number to total  baryon number.
The bulk free energies of heavy nuclei are represented as
\begin{equation} \label{eq:bulk}
F_{i}^{B}(n_{eqi},T) = A_i \omega(n_{eqi},Z_i/A_i,T)    , 
\end{equation}
where $n_{eqi}$ is the equilibrium density of  heavy nucleus $i$ to be optimized and 
 $A_{i}$ is the mass number of nucleus $i$.
The bulk  free energy consists of interaction and  kinetic  terms as follows:
\begin{eqnarray}%
 \label{eq:para1}
 \omega(n_B,x,T) &= & \omega_{int}(n_B,x)  + \omega_{kin}(n_B,x,T) .
\end{eqnarray}
The interaction term is calculated based on Oyamatsu et al. \cite{oyamatsu03} as
\begin{eqnarray}%
 \label{eq:paraint}
\omega_{int}(n_B,x) &=&  4 x(1-x) v_s(n_B) /n_B + (1-2 x)^2 v_n(n_B) /n_B  ,\\
v_s(n_B) & = &a_1 n_B^2 + \frac{a_2 n_B^3}{1+a_3 n_B}   , \\
v_n(n_B) & = &b_1 n_B^2 + \frac{b_2 n_B^3}{1+b_3 n_B},
\end{eqnarray}
where  $v_s$ and $v_n$ are energy densities for symmetric and pure-neutron matter, respectively, and $b_3=$1.58632~fm$^3$.  
The kinetic term is derived from the Fermi integrals of nucleons \cite{lattimer91}:
\begin{eqnarray}%
 \label{eq:parakine}
\omega_{kin}(n_B,x,T) &=& \frac{T}{2 \pi^2 n_B}  \left(\frac{2m_pT}{\hbar^2} \right)^{3/2}  F_{3/2} (\eta_p) +   \frac{T}{2 \pi^2 n_B}  \left( \frac{2m_nT}{\hbar^2} \right)^{3/2}  F_{3/2} (\eta_n)  ,
\end{eqnarray}
where $\eta_{p/n}=\left( \mu^0_{p/n} - \partial \omega_{int}/ \partial n^0_{p/n} \right)/T $ and
$m_p$ and $m_n$ are proton and neutron masses, respectively.
The chemical potentials, $\mu^0_{p/n}$, are derived from the number densities of protons, $n^0_p=x n_B$,  and neutrons, $n^0_n=(1-x) n_B$,   
and $F_{k}(\eta)$  is defined as 
\begin{eqnarray}%
 \label{eq:gfd}
F_{k}(\eta)=\int_0^{\infty} u^{k} \left[ 1+\exp{(u-\eta)} \right]^{-1} {\rm d} u .
\end{eqnarray}

In the  previous model,
the bulk energy per baryon was estimated by the simplified formula: 
\begin{equation}
 \omega(n_B,x,T)= \omega_0 + \frac{K_0}{18 n_0^2} (n_B-n_0)^{2}   + \left[S_0 +\frac{L}{3 n_0}  (n_B-n_0)   \right] (1-2x)^2  - \frac{T^2}{\epsilon_0}  , 
 \label{eq:para0}
\end{equation}
with 
 $\epsilon_0=16$~MeV
\cite{lattimer91,bondorf95}.
The parameters for bulk properties  at zero temperature,  $\omega_0$, $n_0$, $K_0$, $S_0$ and $L$,
characterize the EOS for uniform nuclear matter and  are  useful to compare them with the canonical values obtained from terrestrial experiments and observations of compact stars \cite{oertel17}.  
In the new model, the five coefficients, $a_1$, $a_2$, $a_3$, $b_1$ and $b_2$, in Eq.~(\ref{eq:paraint}) are determined to reproduce those bulk properties in Eq.~(\ref{eq:para0})
at $T=0$ and $n_B=n_0$  \cite{oyamatsu03}: 
\begin{eqnarray}%
\label{eqcoef1}
S_0 &=& \frac{\hbar^2}{12} \left( \frac{3 \pi^2}{2} \right)^{2/3} \left( \frac{1}{m_n}+\frac{1}{m_p} \right)  n_0^{2/3} + (b_1-a_1) n_0 + \left( \frac{b_2}{1+b_3 n_0}  - \frac{a_2}{1+a_3 n_0}  \right) n_0^2  , \\
\label{eqcoef2}
\frac{1}{3}n_0 L &=& \frac{\hbar^2}{18} \left( \frac{3 \pi^2}{2} \right)^{2/3} \left( \frac{1}{m_n}+\frac{1}{m_p} \right)  n_0^{5/3} + (b_1-a_1) n_0^2 + 2 \left( \frac{b_2}{1+b_3 n_0}  - \frac{a_2}{1+a_3 n_0}  \right) n_0^3  \\
& & -\left[ \left( \frac{b_2 b_3}{(1+b_3 n_0)^2} -  \frac{a_2 a_3}{(1+a_3 n_0)^2}  \right) n_0^4  \right] 
\nonumber , \\
\label{eqcoef3}
\omega_0  &=& \frac{3 \hbar^2}{20}  \left( \frac{3 \pi^2}{2} \right)^{2/3} \left( \frac{1}{m_n}+\frac{1}{m_p} \right)  n_0^{2/3}  + a_1 n_0 +  \frac{a_2 n_0^2}{1+a_3 n_0} , \\
\label{eqcoef4}
K_0  &=& \frac{3 \hbar^2}{10}  \left( \frac{3 \pi^2}{2} \right)^{2/3} \left( \frac{1}{m_n}+\frac{1}{m_p} \right)  n_0^{2/3}  +  \frac{18 a_2 n_0^2}{(1+a_3 n_0)^3} , \\
\label{eqcoef5}
0  &=& \frac{\hbar^2}{10}  \left( \frac{3 \pi^2}{2} \right)^{2/3} \left( \frac{1}{m_n}+\frac{1}{m_p} \right)  n_0^{-1/3}  + a_1  +  \frac{2 a_2 n_0}{1+a_3 n_0} - \frac{a_2 a_3 n_0^2}{(1+a_3 n_0)^2}   .
\end{eqnarray}
The last equation derives from the fact that pressure becomes zero at the saturation density.  
The bulk parameters are taken from the  same reference \cite{oyamatsu03} and 
corresponding coefficients given by Eqs.~(\ref{eqcoef1}-\ref{eqcoef5})  are summarized in the Tab.~\ref{tab1_bulk}.
The nuclear bulk energies are shown  in Fig.~\ref{fig_blk}.
The left panel indicates that
the bulk energies for symmetric nuclear matter ($x=0.5$) of the two parameter sets
 are identical, since the difference in  the values of $L$ is not influential at all.
On the other hand, the free energies for asymmetric nuclear matter ($x=0.1$) rise more rapidly for the parameter set B with the larger value of $L$ as shown in the right panel.

{
{
The Coulomb energy of heavy nuclei is expressed in the same way as in the previous model:
\begin{eqnarray}
F_i^C(n_{eqi},n'_p,n_e)= \frac{3}{5} \left(\frac{3}{4 \pi} \right)^{-1/3}  e^2 n_{eqi}^2 \left(\frac{Z_i - n'_p v_{Ni}}{A_i}\right)^2  {v_{Ni}}^{5/3} D(u_i) ,  \label{eqcl} 
\end{eqnarray}  
where $n_e=Y_p n_B$  is  the number density of electrons and  the local number densities of free nucleons are defined as $n'_{p/n}=n_{p/n}/\xi$
with their number densities in the whole volume, $n_{p/n}$, and  the excluded volume effect, $\xi$.
They are expressed as  
$n_{p/n}=N_{p/n}/V$ and $\xi=1-V_N/V$ with the total volume $V$, the nuclear volume $V_N=V(\sum_i n_i v_{Ni}+\sum_j n_j v_{Nj})$ and the numbers of free protons, $N_p$, and free neutrons, $N_n$, in the same way as in the HS and FYSS EOSs.} 
The individual nuclear volumes are estimated as $v_{Ni}=A_i/n_{eqi}$ and $v_{Nj}=A_j/n_0$.} 
Equation~(\ref{eqcl})  is 
obtained within the Wigner-Seitz approximation assuming 
that whole system is divided into electrically-neutral spherical cells each containing one nucleus and corresponding numbers of free nucleons and electrons, see Refs. \cite{furusawa11,ebel18} for details.
The cell volume, volume fraction and its function are expressed as $v_i = (Z_i - n'_p  v_{Ni})/(n_e-n'_p)$, $u_i=v_{Ni}/v_i$ and $D(u_i)=1-\frac{3}{2}u_i^{1/3}+\frac{1}{2}u_i$ and $e$ is the elementary charge. 

The surface energy of heavy nuclei is calculated as
\begin{eqnarray}
F_i^{S} (n_{eqi},n'_p,n'_n,T) = 4 \pi {r^2_{Ni}} \, \sigma_i \left(1-\displaystyle{\frac{n'_p+n'_n}{n_{eqi}}} \right)^2 \left( \displaystyle{\frac{T^2_c(x)-T^2}{T^2_c(x)+T^2}} \right)^{5/4} ,   \label{eqsf}
\end{eqnarray}  
where $r_{Ni} = ( 3/4 \pi v^{Ni})^{1/3}$  is the radius of the nucleus.
We employ
the following surface tension  \cite{agrawal14b} :
$\sigma_{i} =\sigma_0 (16 + C_s) /\left( (1-Z_i/A_i)^{-3} + (Z_i/A_i)^{-3}  +C_s \right)$,  
where $\sigma_0=1.14$ MeV/fm$^2$ and  $C_s =$ 12.1 MeV.
The  critical  temperature $T_c(x=Z_i/A_i)$ is determined to be consistent with the free energy density of bulk nuclear matter described by Eq.~(\ref{eq:para1}). 
It is  defined as the temperature  at which simultaneously $(\partial P_{bulk}/\partial n_B)|_{x}=0$ and $(\partial^2 P_{bulk}/\partial n_B^2)|_{x}=0$,
where the bulk pressure is defined as $P_{bulk}=n_B^2 \partial \omega(n_B,x,T)/\partial n_B$. In the previous model, we just assumed $T_c=18$ MeV for all nuclei.
 Figure~\ref{fig_crt} displays the critical temperatures  above which the bulk nuclear matter can not be bound.
The neutron-rich matter has smaller critical temperatures due to the symmetry energies.
The extremely neutron-rich matter with $x \lesssim 0.1$ has no local minimum point in the bulk pressure. 
It can be confirmed that the larger value of $L$ for the parameter set B  causes slightly lower $T_c$.

The masses of  light clusters, $M_j$,  are set to be their experimental values, $M_j^{data} $, plus the Coulomb energy shifts, which are defined as $\Delta F^C_{j}=F_j^C(n_0,n'_p,n_e)-F_j^C(n_0,0,0) $ by using the same formula for the Coulomb energies  of heavy nuclei, Eq.~(\ref{eqcl}). 
Note that the equilibrium densities of light clusters are assumed to be  constant and equal to the saturation density of symmetric nuclear matter, $n_0$.
In the previous paper,  we represented the light clusters by $\alpha$ particles only, without the Coulomb correction,  for the fair comparison with SNA EOSs.
{
Effective repulsive interactions among light clusters and other baryons are introduced by using excluded-volume effects, as explained in the next paragraph.
Explicit nuclear interactions for light clusters, however, are not considered in this model for simplicity, since there are many ambiguities in the model for the free energy of light clusters and since the main focus of this work is the calculation of heavy nuclei.
We discuss some other approaches for light clusters  briefly in Section~\ref{sec:res2}.}

The translational free energies of light and heavy nuclei in Eq.~(\ref{total}) are
expressed as 
\begin{equation}%
\label{eq:tra}
 F_{i/j}^{t} = T \left\{ \ln \left(\frac{ n_{i/j}/\kappa}{g_{i/j}^0 (M_{i/j} T/2\pi \hbar ^2 )^{3/2}  }\right)- 1 \right\} , 
\end{equation}
where  
$g_{i/j}^0$  are the spin degeneracy factors of the ground state, which are set to unity  for heavy nuclei and  to the experimental values for light clusters.
{
We consider contributions of excited states of heavy nuclei are included in the temperature-dependent mass free energy, $M_i(T)$, in the same way as in SMSM  and FYSS EOSs, 
whereas some MNA EOSs such as HS EOS 
 employ  the partition functions $g_i(T)$ counting the excited state  (see Refs. \cite{buyukcizmeci13,furusawa13a} about the difference between the two approaches).
Excited states for light clusters are also neglected here for simplicity (see Section \ref{sec:res2}).
 The excluded volume correction is employed with $\kappa=1-n_B/n_0$
in the same way as in the SMSM and HS EOSs.
Note that  $\kappa$  differs from the excluded volume effect for free nucleons, $\xi$,
in which only nuclei contribute to the exclusion \cite{hempel10}.} 

{
The free energy densities of 
free nucleons dripped from nuclei
are 
based on the same calculations for 
the bulk energies of heavy nuclei 
with the excluded volume effect. 
It  is expressed  as 
\begin{equation} 
\label{eq:pn}
f_{p/n}=\xi n'_{p/n} \omega(n'_p+n'_n, n'_p/(n'_p+n'_n),T).
\end{equation}
 The free energy per nucleons, $\omega(n_B,x,T)$, is estimated by Eq~(\ref{eq:para1}), and includes kinetic and interaction terms.}  
In the previous model, it was estimated by the translational energies of point-like particles based on the ideal-gas approximation with the  excluded volume.

To optimize  the total free-energy densities, we solve four variables, the chemical potentials and local number densities of nucleons, $n'_{p/n}$, to satisfy four equations,
 the relations between the chemical potentials and number densities of nucleons as well as the baryon- and charge-number conservations.
This process  to solve the full nuclear ensemble is almost the same as the optimization in the MNA EOSs.  
In addition, we find the equilibrium  densities of all nuclei, $n_{eqi}$, by minimizing their  mass free energies, $M_i$,  at every step of the above process.
This equilibration  is usually included in the optimization for SNA EOSs, in which the equilibrium density of the representative nucleus is adjusted.

\section{Result of self-consistent model  \label{sec:res1}}
We discuss the temperature dependence of the results of the EOSs with the bulk parameter sets B and E  for $n_B$=~0.1~$n_0$ and  0.3~$n_0$  and  $Y_p=$ 0.2 and  0.4, typical values in the supernova core, where heavy nuclei prevail even at high temperatures.
 Figure~\ref{fig_dis2}  displays the mass fraction $X_i=A_i n_i/n_B$ and equilibrium  densities, $n_{eqi}$, in the $(N,Z)$ plane for the model with the parameter set E at
$Y_p=0.2$, $n_B=0.3 \ n_0$ and $T=1$, 3, 5 and 10~MeV. 
Generally speaking, the abundant nuclei are determined  by the balance among the chemical potentials,  binding energies and entropies. 
As temperature increases,  the difference in entropies becomes more important than that in binding energies. 
At extremely-high temperature, the neutron-rich nuclei, however, can not survive due to the vaporizations even with large chemical potentials and entropies.
For instance, the nuclei of $(N_i,Z_i)=(60, 20)$  and $(30, 10)$ are populated at  $T=3$ and 5 MeV  but disappear  at $T=$ 10~MeV, albeit relatively close to the abundance peak  $(N_i,Z_i) \sim (18, 6)$. 
Note that the nuclear  equilibration  depends not only on bulk properties but also on Coulomb and  surface energies. 
Those nuclei with $Z_i/A_i=0.25$ 
are diminished  at $T=10$~MeV 
although the temperature is lower than  the critical temperature, $T_c \sim14$ MeV, of the bulk matter with $x=Z_i/A_i=0.25$.
This is in part due to the reductions of the surface tensions  by the factor, $\{ (T^2_c-T^2)/(T^2_c+T^2) \}^{5/4}$.
The vaporizations of neutron-rich nuclei  can be confirmed in Fig.~\ref{fig_iso},
 in which mass fractions  and equilibrium densities of nuclei with $A_i=50$ are displayed
for the models with parameter sets B and E.
The neutron-rich nuclei have smaller equilibrium densities than symmetric nuclei and are cut off one by one  at high temperatures.
We found that the EOS with the parameter set B, which has the larger value of $L$,  gives smaller equilibrium densities. 

Figure~\ref{fig_satu} displays the equilibrium densities of the averaged values, 
 $\langle n_{eq}\rangle=(\sum n_{eqi} n_i)/\sum n_i$, and of the specific nucleus with $(N_i,Z_i)=(30,20)$, $^{50}$Ca, as functions of temperature. 
We find that the equilibrium densities of individual nuclei decrease as temperature increases.
This is because of  the increase of kinetic term in bulk energy, $\omega_{kin}$, as shown in Fig.~\ref{fig_blk} and the decrease of  surface energy.
On the other hand, the average equilibrium density increases around $T=3$~MeV, since the lighter nuclei with $A_i \lesssim100$ take the
 place of the heavy nuclei with $A_i \gtrsim100$ as will  be shown in Fig.~\ref{fig_mamz}.
 The  nuclei with small mass numbers generally have small Coulomb energies per baryon and large  equilibrium densities. 
This inconsistency in the temperature dependence between the average and individual equilibrium densities is similar to that in the density dependence observed in the previous work.
We showed that the individual nuclei are compressed  when few nucleons are dripped, while the average equilibrium density always decreases along with the density rise
because of the change in dominant nuclear species
  \cite{furusawa17c}.
At  temperatures higher than $T\sim5$~MeV, the dominant nuclear species are not greatly changed  
and  the average equilibrium density is reduced according to their decompression.
{
The nuclei with low $Z_i/A_i$, however, start to be vaporized one by one above $T \sim 5$~MeV as shown in Fig. \ref{fig_dis2}. 
As a result, the equilibrium densities increase a little  
for $Y_p=0.2$ when the dominant neutron-rich nuclei  vaporize, 
which have smaller equilibrium densities than symmetric nuclei 
 in general.
For instance, the nuclei with $Z_i/A_i \sim0.40$ such as  $^{50}$Ca are diminished around $T=13$ MeV and $n_B=0.1 \ n_0$  
and, then, the average equilibrium densities change non-smoothly.

Figure~\ref{fig_xfra} provides the mass fractions of dripped nucleons, light clusters ($Z_j \leq 5$ or $N_j \leq 5$ ) and heavy nuclei ($Z_i \geq 6$ and $N_i \geq 6$)  as functions of temperature. 
At high temperatures, the mass fraction of heavy nuclei is reduced,
 since the particles with larger mass numbers give smaller entropies per baryon. 
The differences in those values between parameter sets B and E for $Y_p=0.4$ are smaller than in the case of $Y_p=0.2$, 
since their bulk energies around $x = 0.5$  are not greatly different as shown in Fig.~\ref{fig_blk}.
As for $Y_p=0.2$, the mass fraction of heavy nuclei  in the parameter set B
 is smaller than in the other model
due to the larger value of $L$, while the free neutron fraction is larger.
{
The average mass and atomic numbers for heavy nuclei  are also reduced at high temperatures to increase the entropy per baryon as shown in Fig. \ref{fig_mamz}.
The smaller equilibrium density for the model with the parameter set B provides larger surface energies and smaller Coulomb energies of nuclei and, as a result,
 the larger  average  mass and proton numbers.

\section{Comparison with other models  \label{sec:res2}}
{
We also perform some calculations to compare different models for free energy of heavy nuclei (Models I-IV).
The models are listed on Table~\ref{tab2_model}.
Model I is the self-consistent model of our new calculation, 
whereas incompressible-liquid drop models are utilized in the other models,
in which the equilibrium densities are set to constant values at zero density and zero temperature as $n_{eqi}(n_e,n'_p,n'_n,T)=n_{eqi}(0,0,0,0)$.
In Model III,  the critical temperature for Eq.~(\ref{eqsf})  is assumed to be $T_c=18$ MeV, while
the charge fraction dependence of $T_c$, which is shown in Fig.~\ref{fig_crt}, is taken into account in Model II.
The former may be regarded as a surrogate for SMSM and FYSS EOSs and 
the latter may be close to a recent MNA EOS, SRO EOS \cite{schneider17}, in which 
 the $Z_i/A_i$ dependence of $T_c$  is introduced in surface tensions.}

{
Recently Pais and Typel \cite{pais16} also construct an MNA EOS with an assumption of the dissolutions of heavy nuclei around $T\sim11$ MeV.
In Model IV, we introduce the same nuclear dissolution factor, $\gamma(T)$, as 
\begin{eqnarray}
n_i&=&\kappa}{g_{i} (M_{i} T/2\pi \hbar ^2 )^{3/2}  {\rm exp} \left( \frac{\mu_i-M_i}{T} \right) \gamma(T) ,  \label{eq:flush} \\
 \gamma(T)&=& \left\{ \begin{array}{ll}
  {\rm exp}\left[ -\left( \frac{T}{T_{fl}-T} \right)^2 \right] \ \ &  {\rm{if}} \   T<  T_{fl},   \\
 0 \ \   &  {\rm if}  \ T> T_{fl},  \label{eq:fla} 
\end{array} \right.
\end{eqnarray}
where $\mu_i$ is the chemical potential of heavy nucleus $i$.     
The flashing temperature, $T_{fl}=$11.26430 MeV, is determined by the local instability of symmetric nuclear matter calculated by the DD2 relativistic mean field model \cite{typel10}.
The value taken from the reference \cite{pais16} is not  consistent with the flashing temperature 
 of our bulk matter calculation.
It is between 12 and 16 MeV as shown in the left panel of Fig.~\ref{fig_blk}, where the local  minimum of  bulk energy around $n_0$ disappears.
In the mass evaluation for Model IV, the temperature dependence of surface tension is ignored or $T_c=\infty$ to avoid the double count of the finite temperature effect.
For all four models, we utilize the same formulations for free energies of the nucleons  with the parameter set B  and light clusters to focus only on modeling heavy nuclei.}

{
 Figures~\ref{fig_xfracmp} displays the mass fraction of heavy nuclei for the four models. Below $T\sim 3$ MeV,  the incompressible models, Models II-IV,  
are not so deviated from the self-consistent model, Model I.
Their mass fractions, however, are underestimated a little up to $T \sim 10-14$~MeV,
since, in Model I, the adjustment of equilibrium densities lowers the mass free energies \cite{furusawa17c}. 
On the other hand, the incompressible  heavy nuclei in Models II and III keep on prevailing even above $T\sim 12-14$~MeV and their fractions are significantly overestimated.
The $Z_i/A_i$ dependence of $T_c$ in  Model II  reduces the surface energies  and increases the mass fraction of neutron-rich nuclei  compared with those of Model III. Such  neutron-rich nuclei, however,  can not survive above $T_c$, and, thus, the mass fraction of the former becomes smaller than that of the latter at $T\gtrsim12$~MeV, $n_B= 0.30 \ n_0$ and $Y_p=0.20$.
The reduction of heavy nuclei in Model IV is quantitatively  similar to that in Model I 
and, thus, the dissolution factor in Model IV is likely to reproduce the nuclear vaporization to some extent even with the incompressible liquid drop model. 
The nuclear dissolution temperatures, $\sim10$ MeV, and the flashing temperature $T_{fl}$  in Model IV are lower than the vaporization  temperatures in Model I, $\sim12-14$ MeV.
These temperatures are, respectively, related to the local instability of symmetric nuclear matter and the global instabilities of compressible liquid nuclei with various asymmetry dependences.
Both the dissolution factor of Eq.~(\ref{eq:flush}) in Model IV and the estimate of surface tensions in Model I, however, are simple and not greatly reliable. Thus, quantitative justifications about the melting temperature would be difficult in this comparison.}

{
Finally we compare some theoretical approaches for light clusters. 
In Models V and VI, the incompressible liquid drop model with $T_c=18$ MeV for heavy nuclei is utilized as in Models III, but we replace the light cluster calculation 
 by other approaches.  
In Model V, a quantum approach is incorporated to evaluate Pauli and Self energy shifts of  light clusters \cite{roepke09,typel10} as  $ M_{j}=M_j^{data}  +\Delta F_j^C+ \Delta F_{j}^{Pa}  + \Delta F_j^{SE}$ in the similar way as in FYSS EOS. 
The Pauli energy shift, $\Delta F_{j}^{Pa}$, is calculated by Eq.~(72) in Typel et al. 2010 \cite{typel10}.  
The self energy shift, $\Delta F_{j}^{SE}$, is evaluated based on interaction energies, $\omega_{int}(n_B,x)$ [Eq.~(\ref{eq:paraint})], for dripped nucleons as 
\begin{eqnarray}
\label{SE}
\Delta E_{j}^{SE}(n'_{p},n'_{n},T) &=& (A_j-Z_j) \Delta E_{n}^{SE}+ Z_j \Delta E_{p}^{SE},  \\
x'&=&n'_p/(n'_{p}+n'_{n}) \\
\Delta E^{SE}_{p}&=&x' \ \omega_{int}(n'_{p}+n'_{n},x') , \\
\Delta E^{SE}_{n}&=&(1-x') \ \omega_{int}(n'_{p}+n'_{n},x') .
\end{eqnarray}
The excluded volume effects in Eq.~(\ref{eq:tra}) are ignored with $\kappa=1$, since Pauli exclusion  is incorporated in the mass evaluation (see ref. \cite{hempel11} about the detailed comparison between  the two approaches).
In the other three models, the masses are evaluated as $ M_{j}=M_j^{data}  +\Delta F_j^C$ as explained in Section~\ref{sec:model}.
Model VI includes the temperature-dependent partition function for light clusters as 
\begin{eqnarray} %
\label{eq:lv}
g_{j}(T) &=& g_j^0 + \frac{0.2}{A_j^{5/3}} \int_0^{16.2A_j} {\rm d} E \ e^{-E/T}  {\rm exp} (\sqrt{2(A_j/8)(1-0.8A_j^{-1/3}) E} , \\
n_j &=& g_j (T) \kappa (M_{j} T/2 \pi \hbar^2 )^{3/2}  {\rm exp} \left( \frac{\mu_j-M_j}{T} \right),
\end{eqnarray}
where $\mu_j$ is  the chemical potential of light cluster $j$. 
This partition function \cite{fai82}  is adopted for all nuclei in HS EOS with the maximum energy modified. 
For the other models, we assume that $g_j(T)=g_j^0$.
The properties of these models are summarized in Table~\ref{tab2_model}.}

{
Figure~\ref{fig_light} displays the mass fractions of deuteron, $X_d$, triton plus helion, $X_t+X_h$, and alpha particle, $X_{\alpha}$, for  Models I, III, V and VI. 
{Model I gives the rapid increase in mass fraction of light clusters at the vaporization point of heavy nuclei, whereas Model III shows the smooth slope around $T\sim$14~MeV. 
The vaporization temperatures of heavy nuclei in the new model  may be checked in the heavy-ion reactions at intermediate energies by measuring  the excitations of light clusters.}
The Pauli energy shifts in  Model V reduce the mass fractions of light clusters compared with those in Model III for the most part. 
Especially for neutron-rich conditions $Y_p=0.2$, their mass fractions are suppressed because of strong Pauli-exclusion among neutrons. Only at $Y_p=0.4$ and for deuteron, the self-energy shift increases the mass fraction a little.
For Model VI, the increase in degree of freedom  at high temperature increases the mass fractions of light clusters.
The model for light clusters also affects the mass fraction of heavy nuclei,  and vice versa. Actually the mass fraction of heavy nuclei  in Models V (VI) tends to be more (less) than that in Model III even with the same free energy model for heavy nuclei.
On the other hand, Fig.~\ref{fig_light} shows that  Model I gives less (more) light clusters than Model III at low (high) temperatures according to the change of mass fraction of heavy nuclei by the self-consistent optimization shown in Fig.~\ref{fig_xfracmp}, 
even if they employ the same model for light clusters.

\section{Conclusion \label{sec:conc}}

The new self-consistent calculation of nuclear abundances and nuclear equilibrium densities 
allows us to investigate how the full ensemble of nuclei behaves under the hot and dense
stellar environment.
We find that the nuclear equilibrium densities decrease  and  finally  the neutron-rich nuclei are dissolved
 to nucleons  one after another  as temperature increases. 
The reduction of surface energies at high temperatures 
and  the increase of kinetic contributions in bulk energies enhance the vaporization,  and finally  lead to disappearance of nuclei even at the temperatures
 below the critical temperatures for bulk matter.
The larger value of the slope parameter, $L$, provides the smaller equilibrium densities and
  lower critical temperatures due to the steeper slope of bulk energies for neutron-rich matter.
The total  mass fractions are reduced and the average mass and atomic numbers are increased owing to the small equilibrium densities in the model with the larger value of $L$. 

These modifications of equilibrium densities and vanishing of hot nuclei are not included in supernova EOSs considering
 the full nuclear ensemble beacuse of the simplifications adopted for nuclear binding energies.
In the ordinary multi-nucleus EOSs,  heavy nuclei are assumed to prevail even at $T \sim 17$ MeV.
We also find that the mass fractions in such incompressible liquid drop models
are underestimated below about $T=10$ MeV or overestimated at higher temperatures due to the neglect of the decompression or vaporization, respectively.
{
We also confirm that  the dissolution factor  \cite{pais16}  may be able to  reproduce the disappearance of heavy nuclei qualitatively even with the incompressible liquid drop model.
In addition,  the mass fractions of light clusters are revealed to be sensitive to their mass evaluations, partition functions and treatment of excluded volume effects
as well as the free energy model for heavy nuclei.}


Our self-consistent calculations including  the optimization for both nuclear abundances and
 structures may not be  suitable for constructing data tables covering a wide rage of
conditions $(n_B,T,Y_p)$,
 because of the huge computational resources required.
{
Tentatively, feasible improvements for  the previous multi-nucleus EOSs,
 in which the critical temperature 
was set to be $T_c=$18 MeV  \cite{buyukcizmeci14,furusawa17a, furusawa17d} or 
 $T_c= \infty$ \cite{hempel10, steiner13} for all nuclei, would be the consideration of  the dissolution factor \cite{pais16}  or  the iso-spin dependence of $T_c$
\cite{schneider17}.}

In our model, the surface tension and its temperature and density dependences are estimated simply to calculate the whole ensemble of nuclei.
  More sophisticated models such as Thomas Fermi calculations would improve results quantitatively  although their self-consistent calculations may be more difficult.  
{
The free energy of light clusters should be also improved as discussed in the last  part of Section~\ref{sec:res2} and
 may influence in a essential way  on the supernova dynamics
 and proto-neutron-star cooling \cite{sumiyoshi08,typel10,furusawa13b}.
More detailed analysis about the light clusters \cite{typel10,hempel11,typel17, avancini17}
 may be required to discuss the properties of  hot nuclear matter in addition to the descriptions of uniform nuclear matter and heavy nuclei.}

\begin{acknowledgments}
S.F.  was supported by Japan Society for the Promotion of
Science Postdoctoral Fellowships for Research Abroad. 
I.M. acknowledges partial support from the Helmholz International Center for FAIR (Germany).
This work was supported by RIKEN iTHES and iTHEMS Projects.
A part of the numerical calculations were carried out on  PC cluster at Center
for Computational Astrophysics, National Astronomical Observatory of Japan.
\end{acknowledgments}

\bibliography{reference180223}

\newpage

\begin{table}[t]
\begin{tabular}{|c||c|c|}
\hline 
\hline
 parameter set  & B & E  \\
 \hline
  $n_{0}$ [fm$^{-3}$]  &  0.15969 &  0.15979 \\
 $\omega_0$ [MeV] &-16.184  &   -16.145 \\
 $K_0$ [MeV] & 230   &  230          \\
 $S_0$ [MeV]  & 33.550 &   31.002  \\
 $L$ [MeV] & 73.214  & 42.498 \\
\hline
$a_1$  [MeV fm$^3$]  & -467.51 & -466.32 \\
$a_2$  [MeV fm$^6$] & 2206.0 & 2192.9  \\
$a_3$ [fm$^3$] & 341.77 & 340.02 \\
$b_1$  [MeV fm$^3$]  & -251.42 & -206.48 \\
$b_2$  [MeV fm$^6$]  & 1138.8 & 662.27  \\
 \hline
\hline
\end{tabular}
\caption{\label{tab1_bulk}%
The parameters of bulk nuclear matter \cite{oyamatsu03,oyamatsu07}.} 
\end{table}

\begin{table}[t]
\begin{tabular}{|c||c|c|c|c|}
\hline 
\hline
 \ & $n_{eqi}$ & $T_c$ & $\gamma(T)$ & light clusters\\
 \hline
Model I & optimized & $T_c(Z_i/A_i)$&  no & excluded volume, $g_j^0$  \\
Model II & constant & $T_c(Z_i/A_i)$ &  no & excluded volume, $g_j^0$  \\
Model III & constant & 18 MeV &  no & excluded volume, $g_j^0$  \\
Model IV & constant & $\infty$ &  yes & excluded volume, $g_j^0$  \\
\hline
Model V & constant & 18 MeV &  no & quantum approach, $g_j^0$  \\
Model VI & constant & 18 MeV &  no & excluded volume, $g_j(T)$  \\
 \hline
\hline
\end{tabular}
\caption{\label{tab2_model}%
Model list for Sec.~\ref{sec:res2}.
The columns of $n_{eqi}$, $T_c$,  $\gamma(T)$, and light clusters
denote respectively whether the nuclear equilibrium density is solved self-consistently, the critical temperature for surface tension in Eq.~(\ref{eqsf}), 
whether the dissolution factor of Eq.~(\ref{eq:fla})  is introduced, and the model free energy density of  light clusters. All the  models adopt the parameter set B for bulk nuclear matter.} 
\end{table}

\begin{figure}
\includegraphics[width=8.1cm]{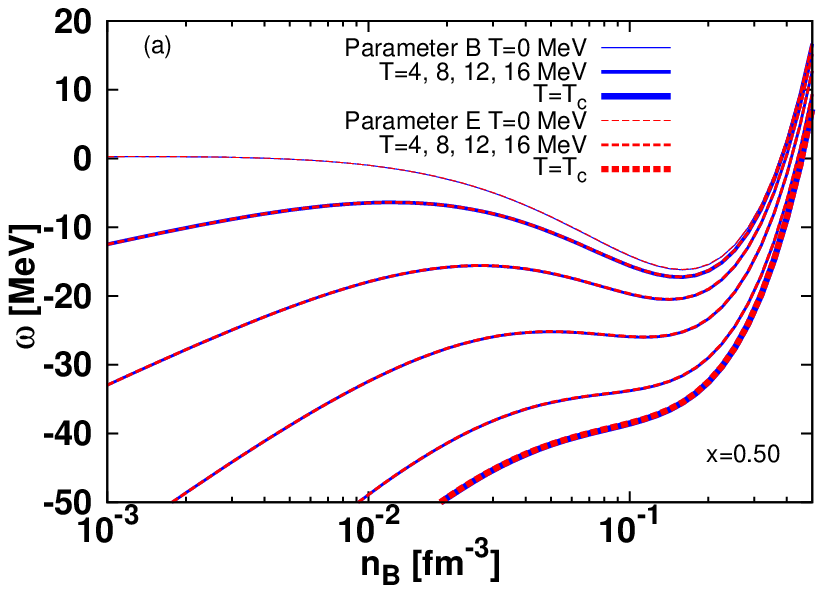}
\includegraphics[width=8.1cm]{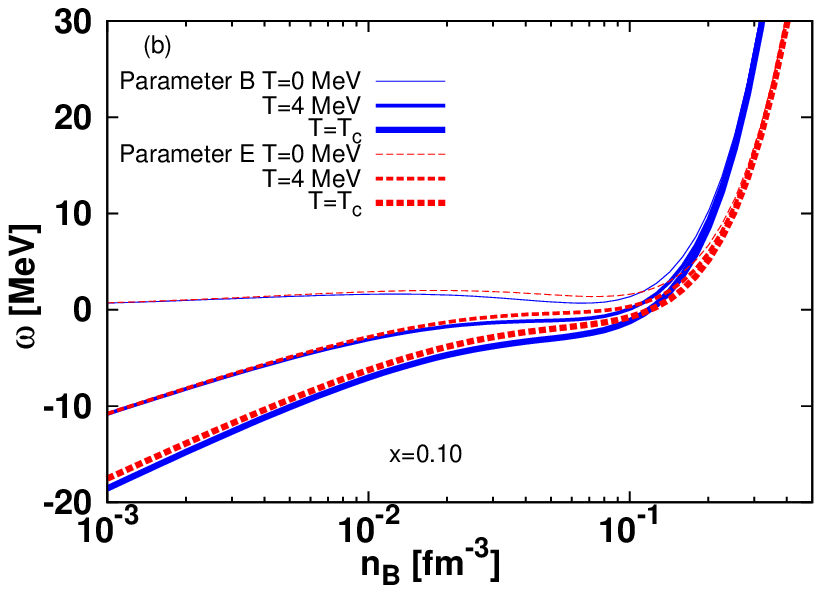}
\caption{Free energies per baryon of  symmetric (left panel, $x=0.5$)  and  asymmetric (right panel, $x=0.1$) uniform nuclear matter for parameter sets  B (blue solid lines) and E (red dashed  lines) at $T=$ 0 (thin lines), 4, 8, 12,  and 16 MeV (medium lines) and $T_c$ (thick lines). The critical temperatures are 18.1 MeV  for both parameter sets at $x=0.5$ and are 5.7 and 5.5 MeV for the parameter sets B and E,  respectively, at $x=0.1$ as shown in Fig.~\ref{fig_crt}.
}
\label{fig_blk}
\end{figure}

\begin{figure}
\includegraphics[width=8.1cm]{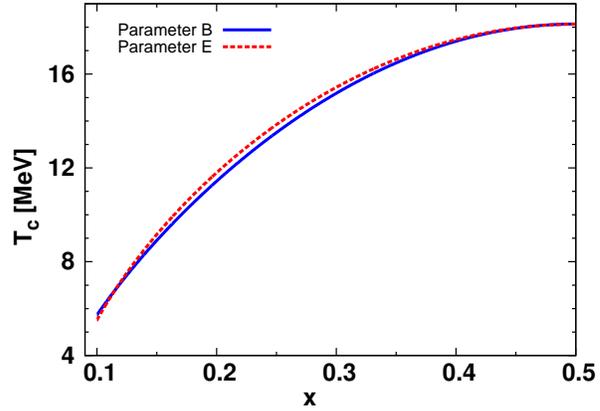}
\caption{The critical temperatures for bulk nuclear matter, above which the bulk pressure has no local minimum,  as a function of charge fraction 
for parameter sets  B (blue solid lines) and E (red dashed  lines).}
\label{fig_crt}
\end{figure}

\begin{figure}
\includegraphics[width=7.5cm]{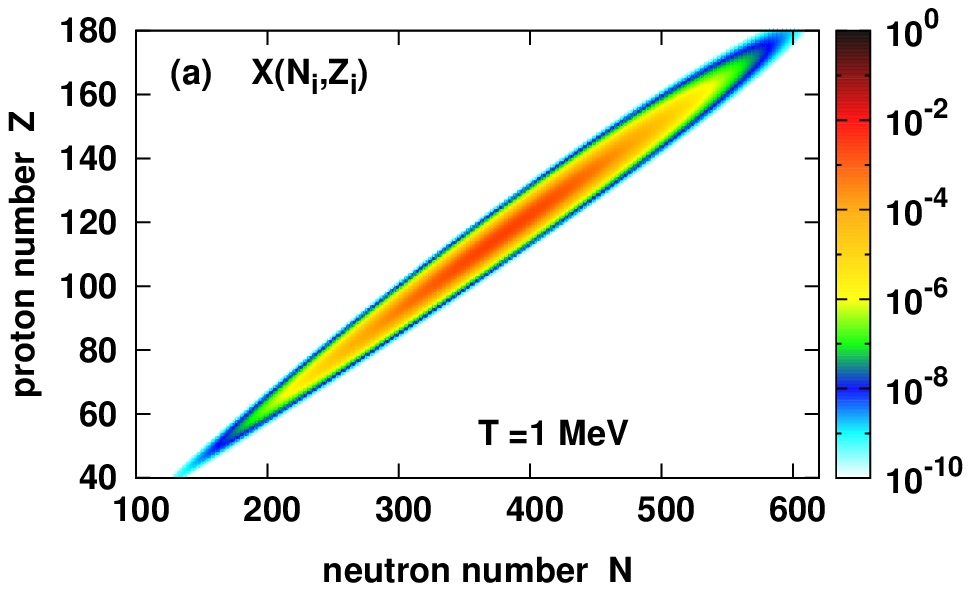}
\includegraphics[width=7.5cm]{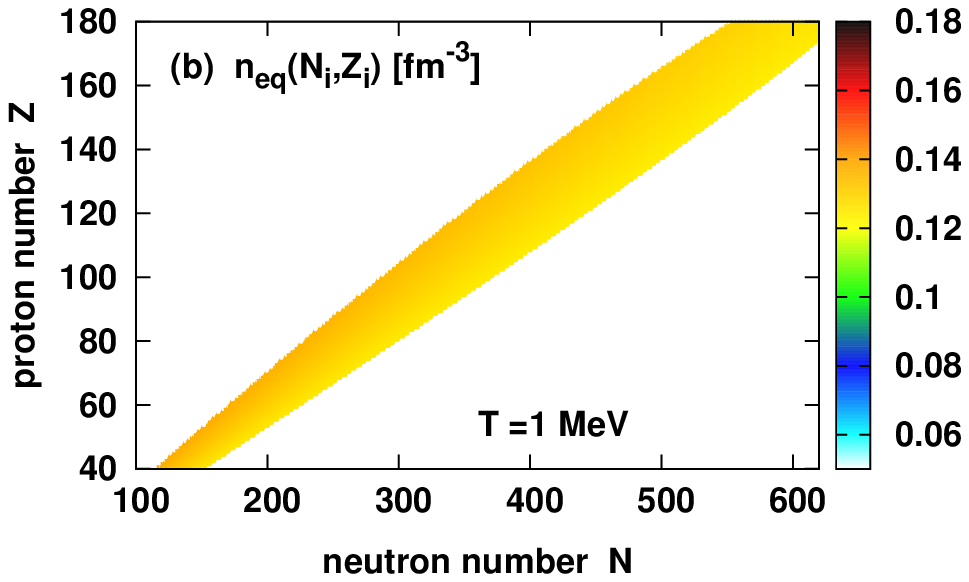}
\includegraphics[width=7.5cm]{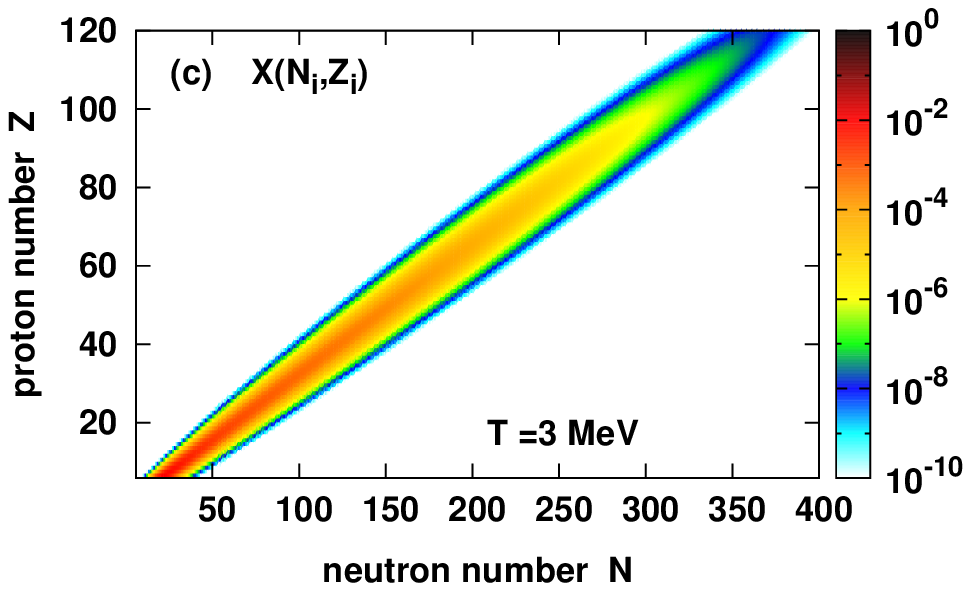}
\includegraphics[width=7.5cm]{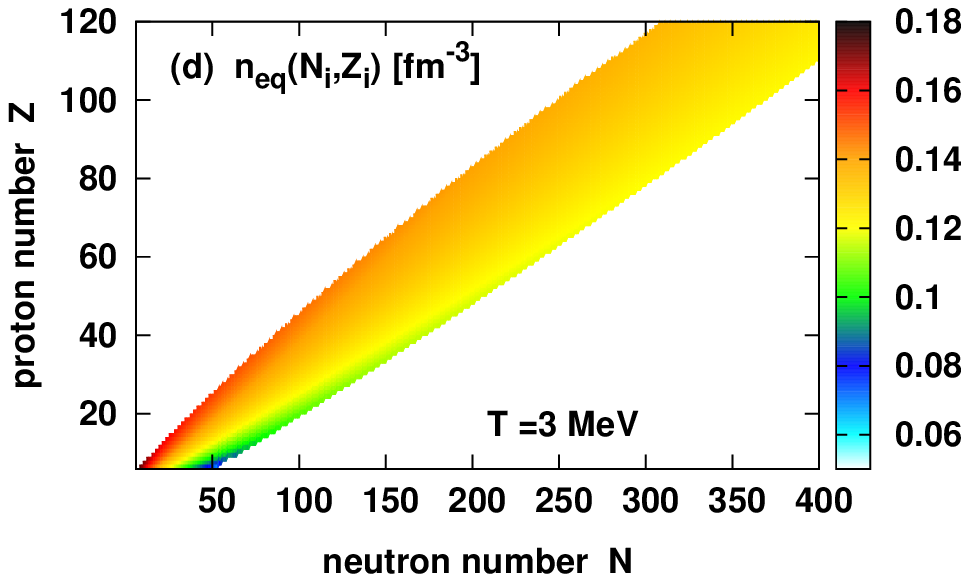}
\includegraphics[width=7.5cm]{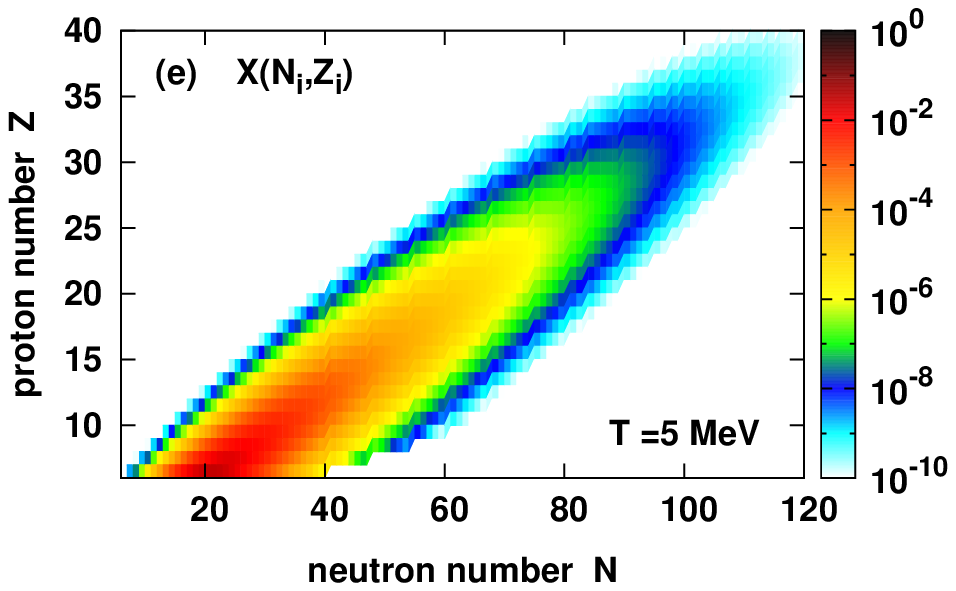}
\includegraphics[width=7.5cm]{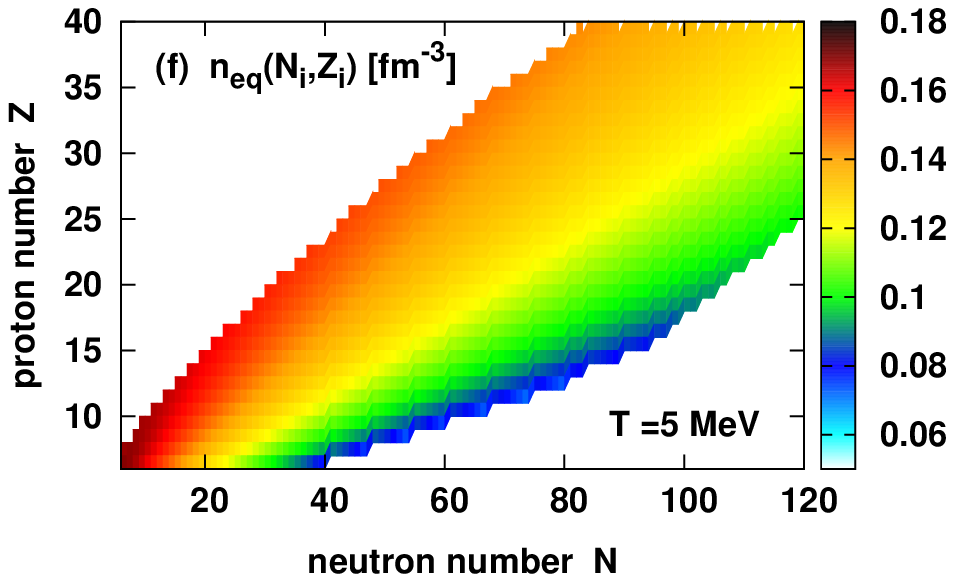}
\includegraphics[width=7.5cm]{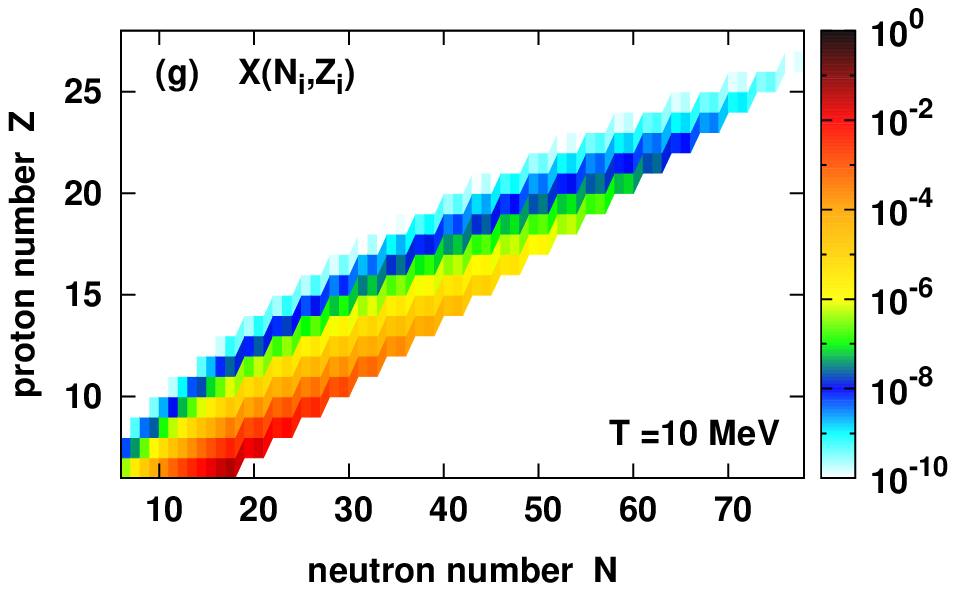}
\includegraphics[width=7.5cm]{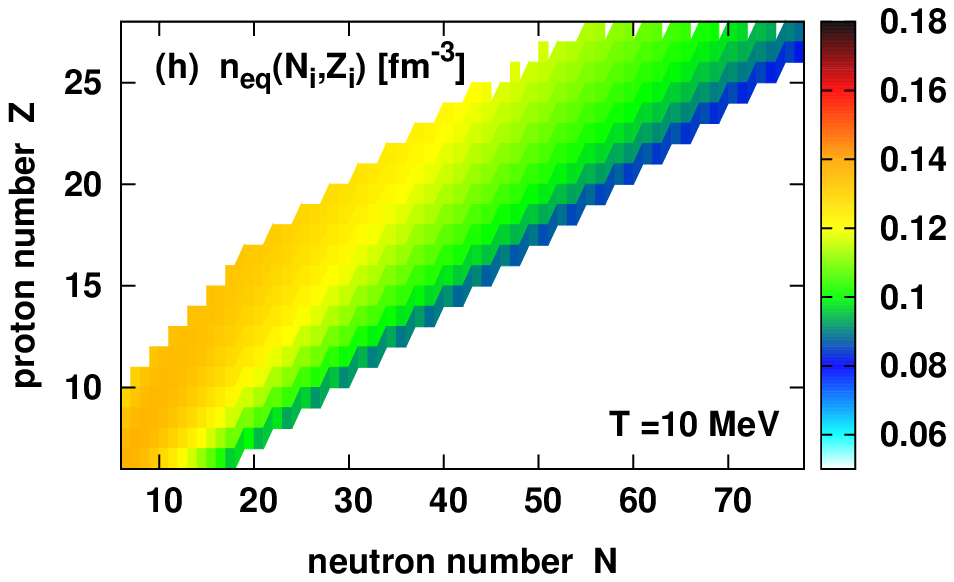}
\caption{
 Mass fraction, $X_i$,  (left column) and equilibrium  density, $n_{eqi}$, (right column)
in $(N, Z)$ plane   for the parameter set  E
at $n_B=0.3$~$n_0$, $Y_p=0.2$, and $T=$~1, 3, 5, and 10  MeV (from top to bottom). 
}
\label{fig_dis2}
\end{figure}

\begin{figure}
\includegraphics[width=7.3cm]{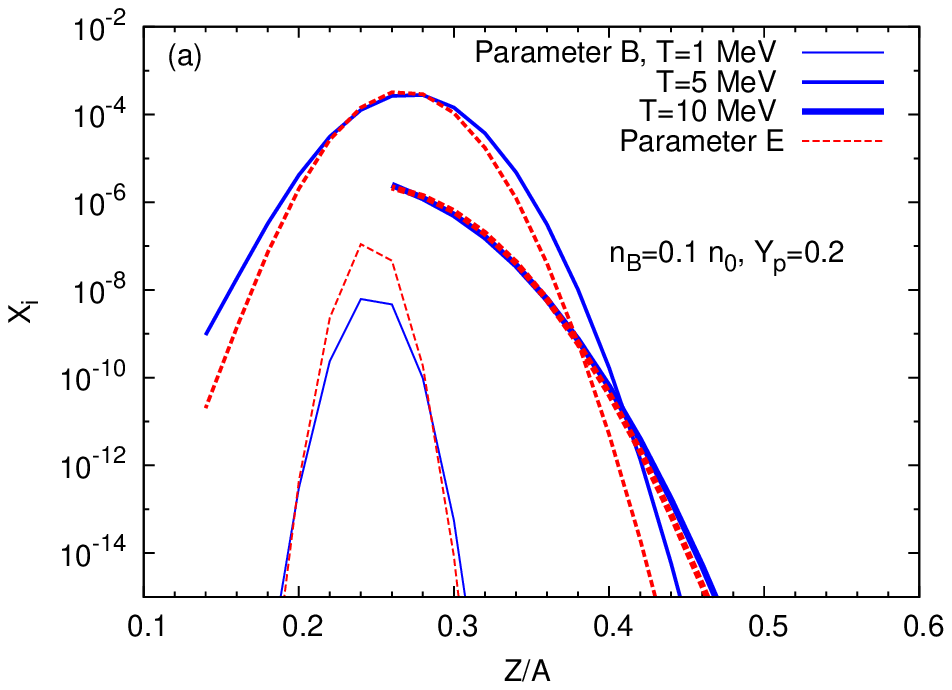}
\includegraphics[width=7.3cm]{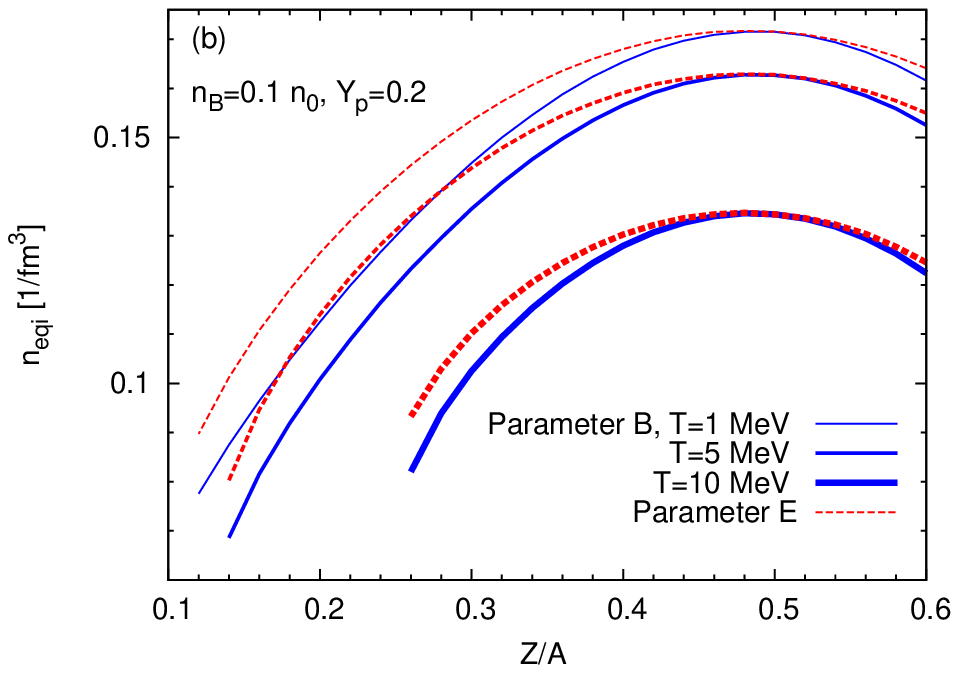}
\includegraphics[width=7.3cm]{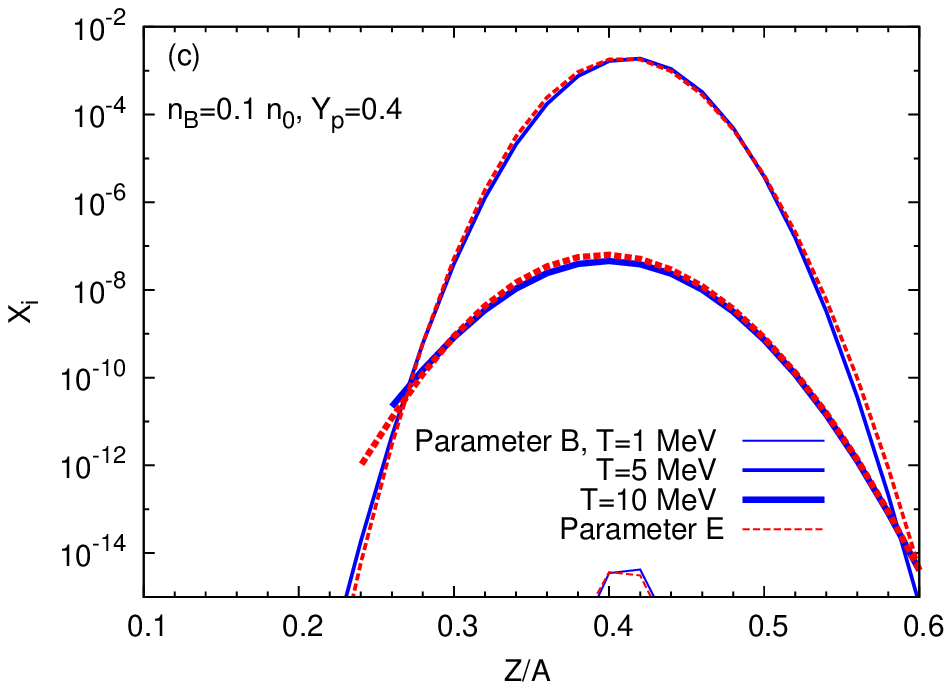}
\includegraphics[width=7.3cm]{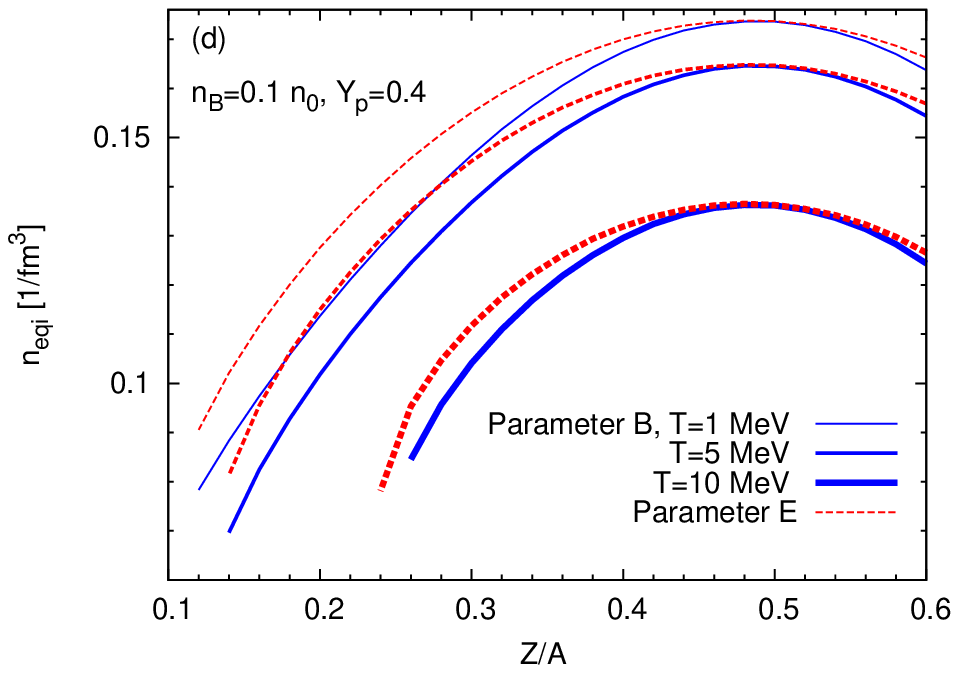}
\includegraphics[width=7.3cm]{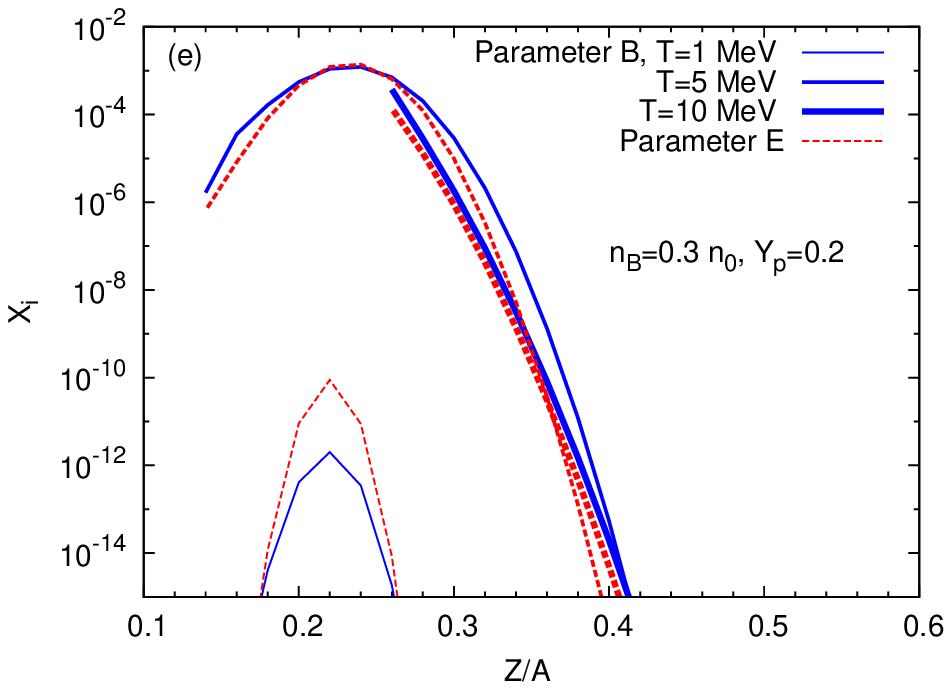}
\includegraphics[width=7.3cm]{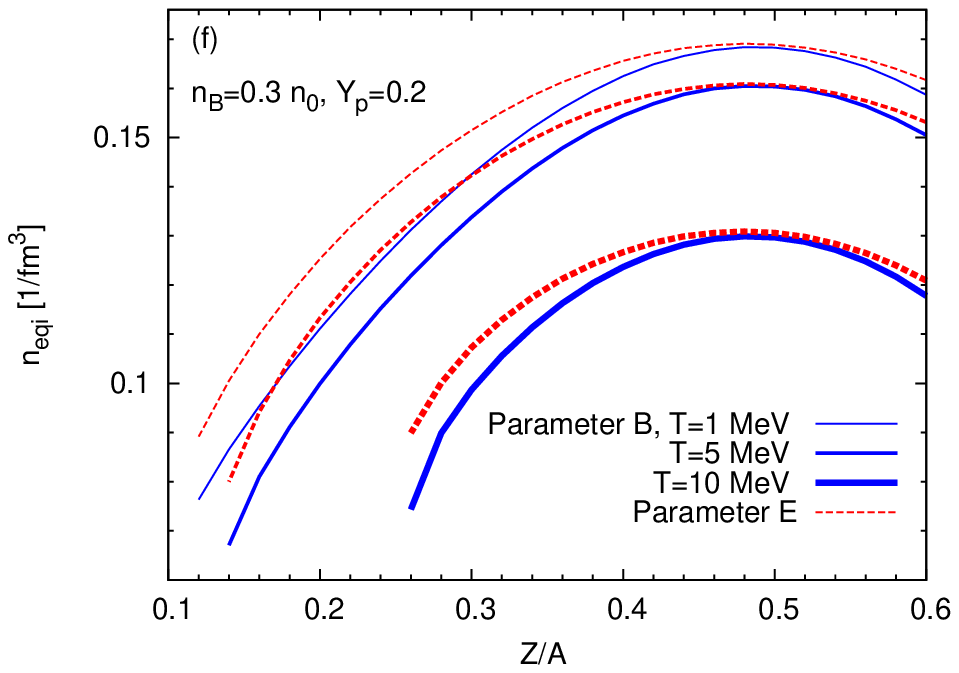}
\includegraphics[width=7.3cm]{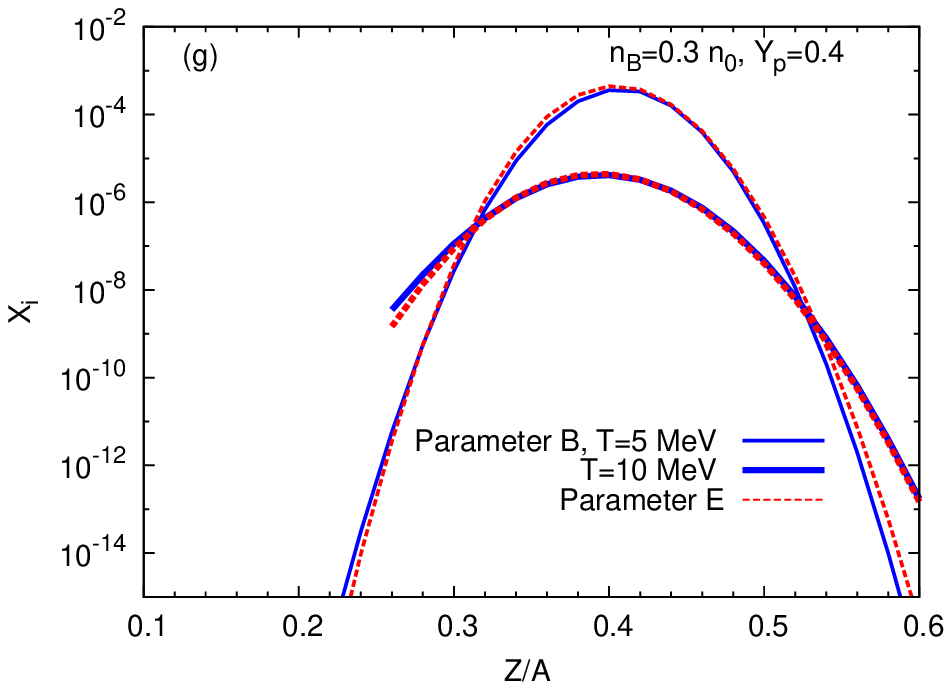}
\includegraphics[width=7.3cm]{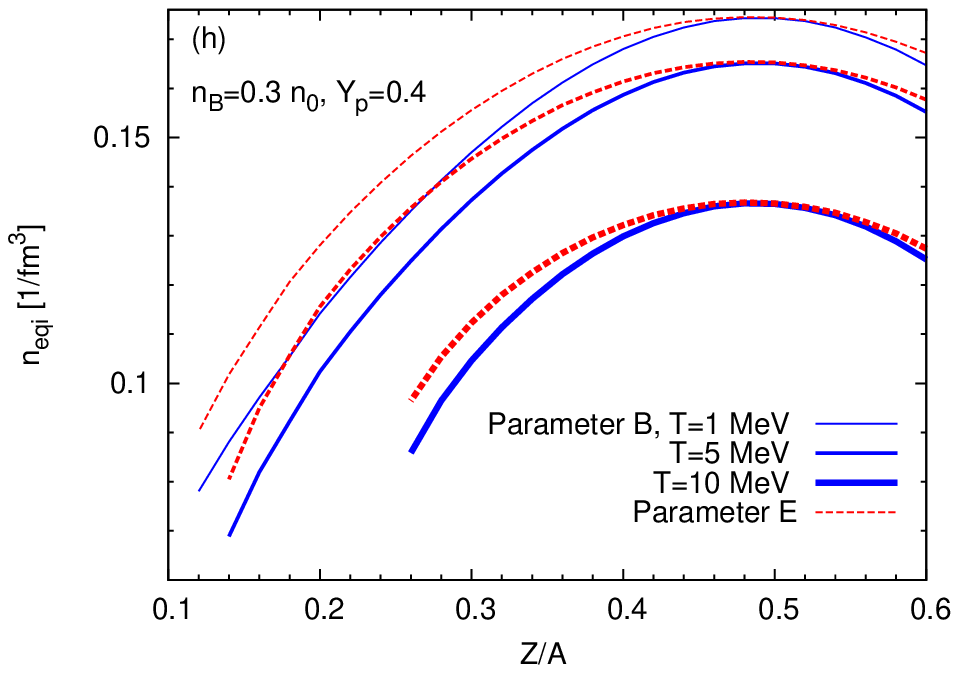}
\caption{
 Mass fraction, $X_i$,  (left column) and equilibrium  density, $n_{eqi}$, (right column)
for the nuclei with $A_i=50$ as functions of charge fraction, $Z_i/A_i$,  
 for parameter sets  B (blue solid lines) and E (red dashed  lines) 
at  $T=$1 MeV (thin lines), 5 MeV (medium lines) and 10  MeV(thick lines)
and at $(n_B,Y_p)= (0.1 \ n_0, \ 0.2)$, $ (0.1 \ n_0, \ 0.4)$, $(0.3 \ n_0, \ 0.2)$, and $ (0.3 \ n_0, \ 0.4)$ from top to bottom.}
\label{fig_iso}
\end{figure}

\begin{figure}
\includegraphics[width=8.1cm]{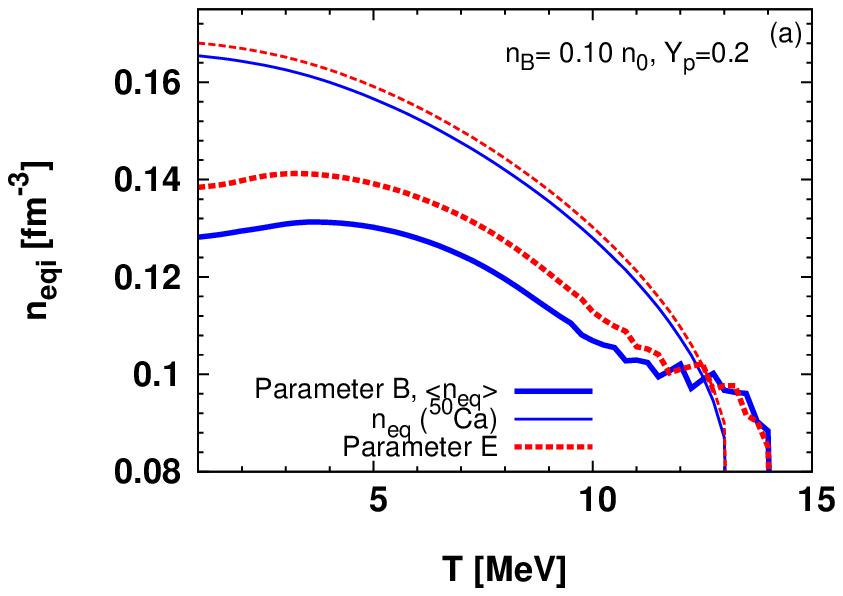}
\includegraphics[width=8.1cm]{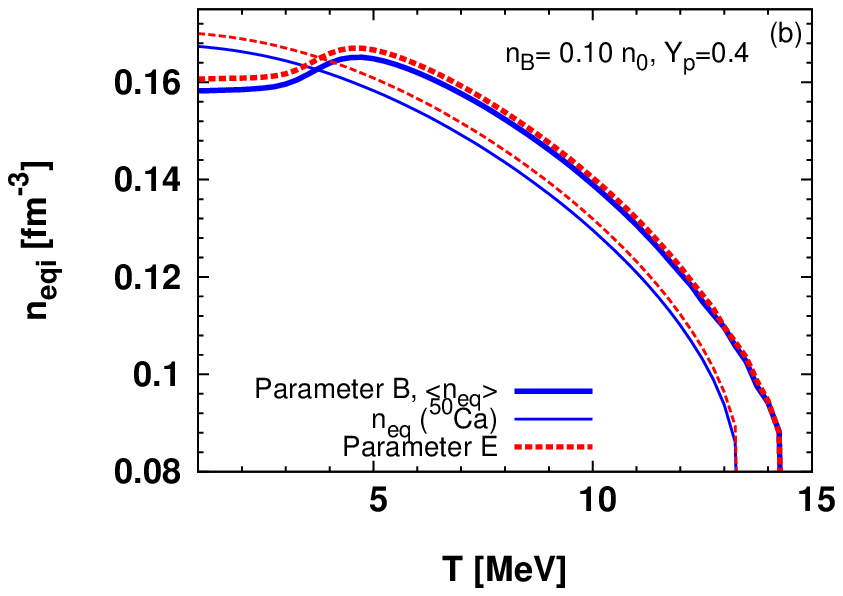}
\includegraphics[width=8.1cm]{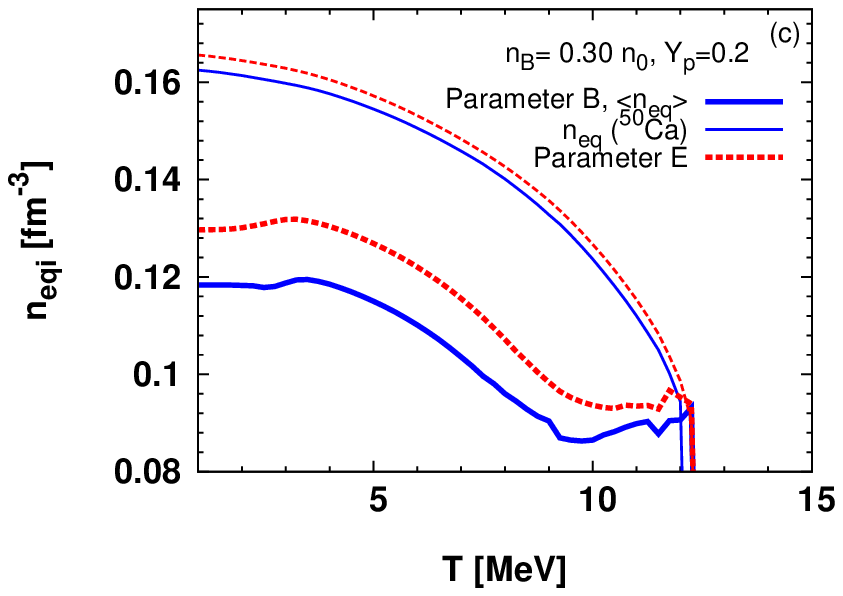}
\includegraphics[width=8.1cm]{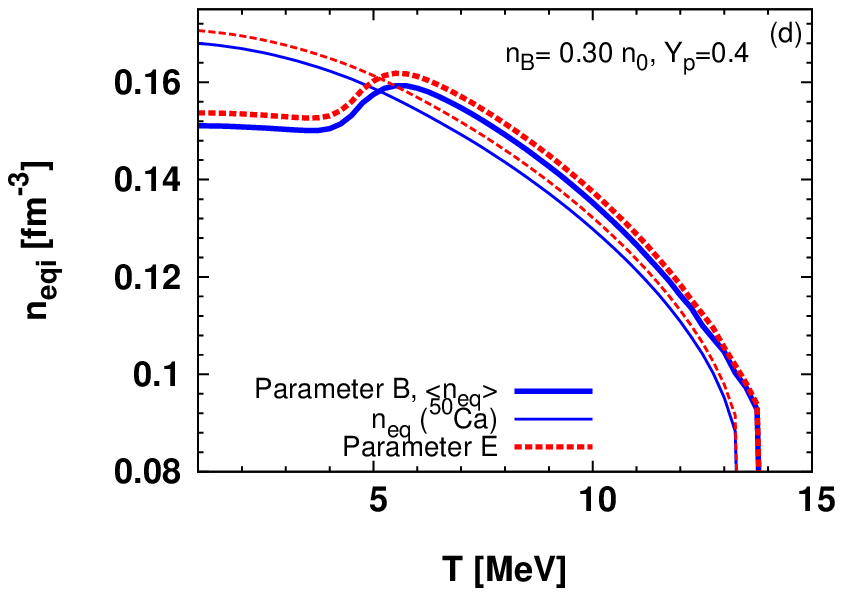}
\caption{Equilibrium density of $^{50}$Ca  (thin lines) and average one over all the heavy nuclei,  $\langle n_{eq} \rangle$, 
(thick lines) as  functions of  $T$
for parameter sets  B (blue solid lines) and E (red dashed  lines) 
at  $n_B=0.1 \ n_0$ (top row) and  $0.3  \ n_0$ (bottom row)
and $Y_p=$ 0.2 (left column) and 0.4 (right column).
}
\label{fig_satu}
\end{figure}

\begin{figure}
\includegraphics[width=8.1cm]{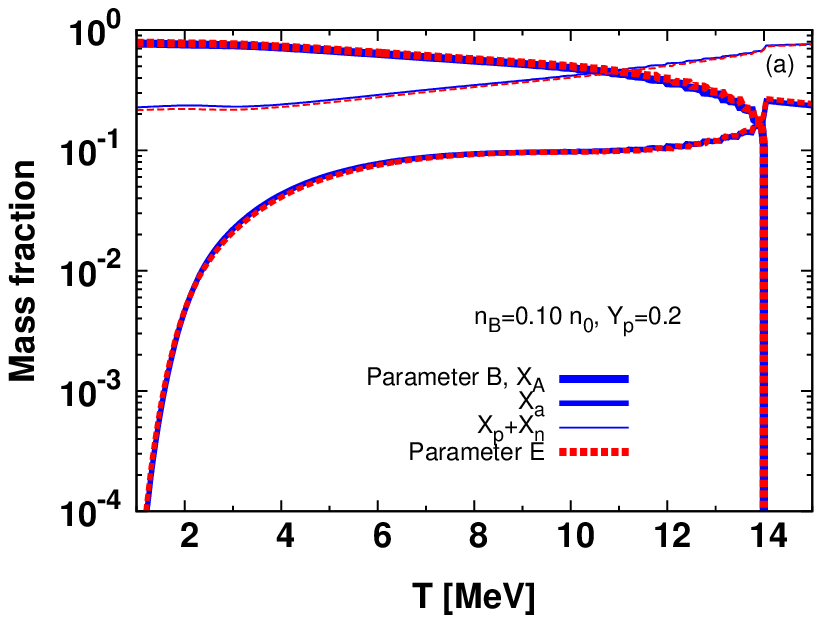}
\includegraphics[width=8.1cm]{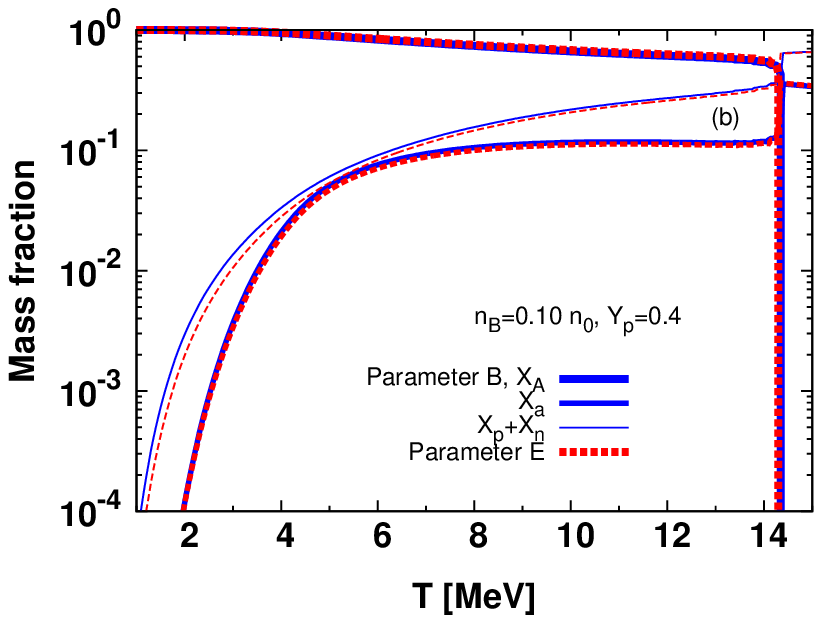}
\includegraphics[width=8.1cm]{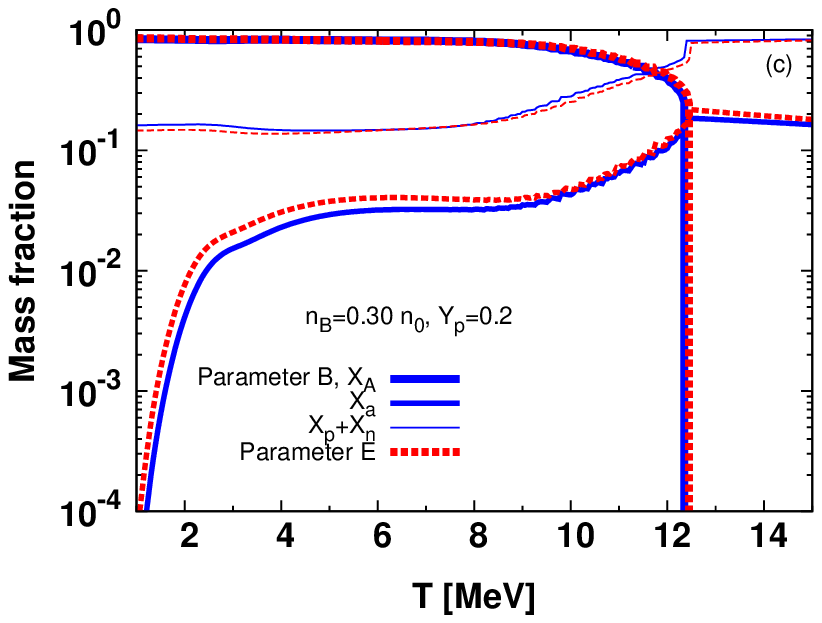}
\includegraphics[width=8.1cm]{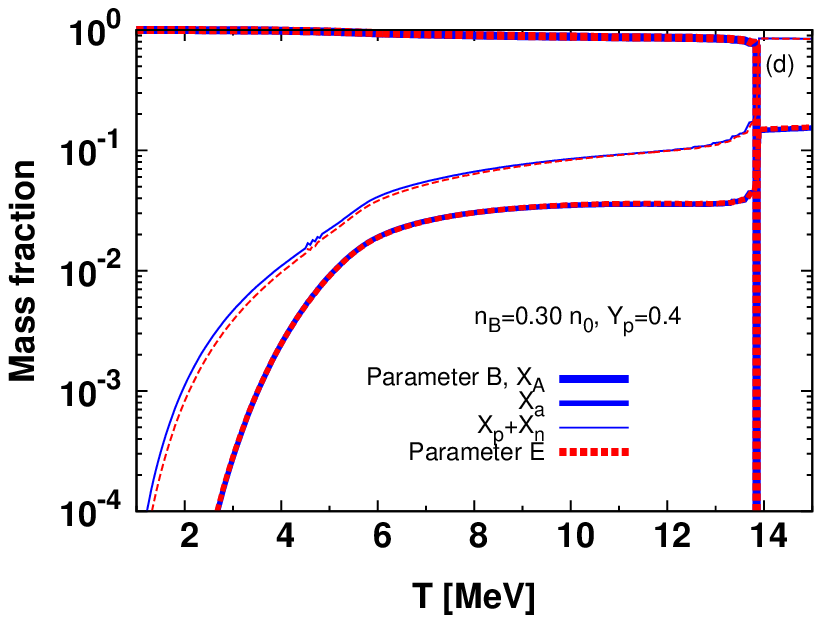}
\caption{Mass fractions of  heavy nuclei (thick  lines), light clusters (medium lines) and free nucleons (thin lines)  as functions of  $T$
for parameter sets  B (blue solid lines) and E (red dashed  lines)
at  $n_B=0.1 \ n_0$ (top row) and  $0.3  \ n_0$ (bottom row)
and $Y_p=$ 0.2 (left column) and 0.4 (right column).
}
\label{fig_xfra}
\end{figure}

\begin{figure}
\includegraphics[width=8.1cm]{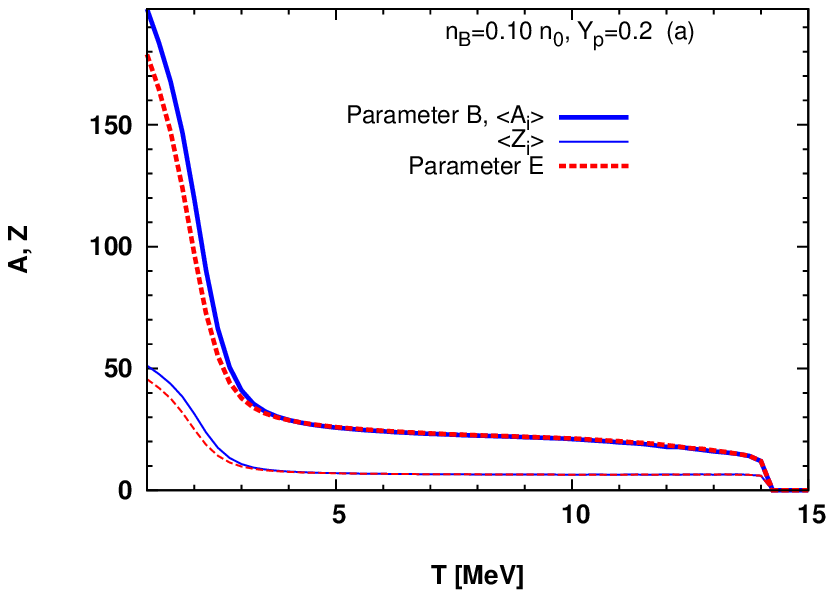}
\includegraphics[width=8.1cm]{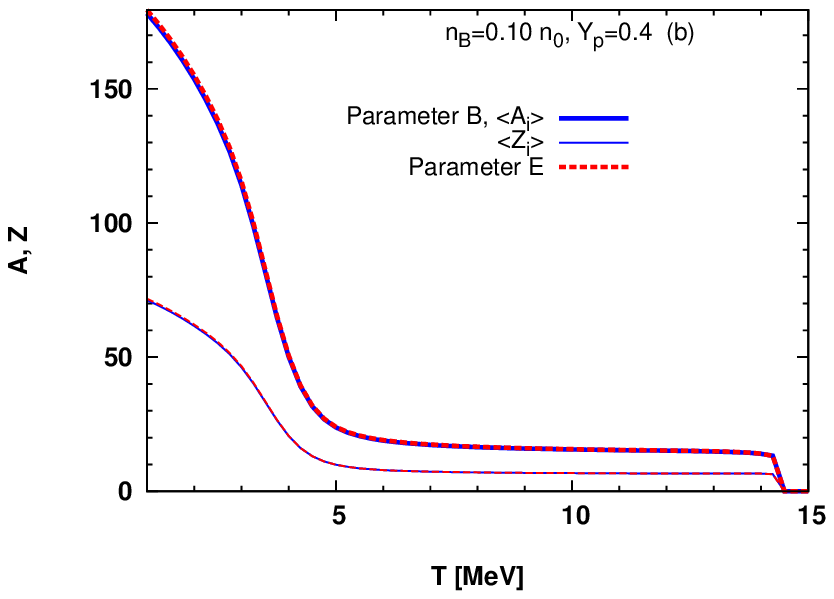}
\includegraphics[width=8.1cm]{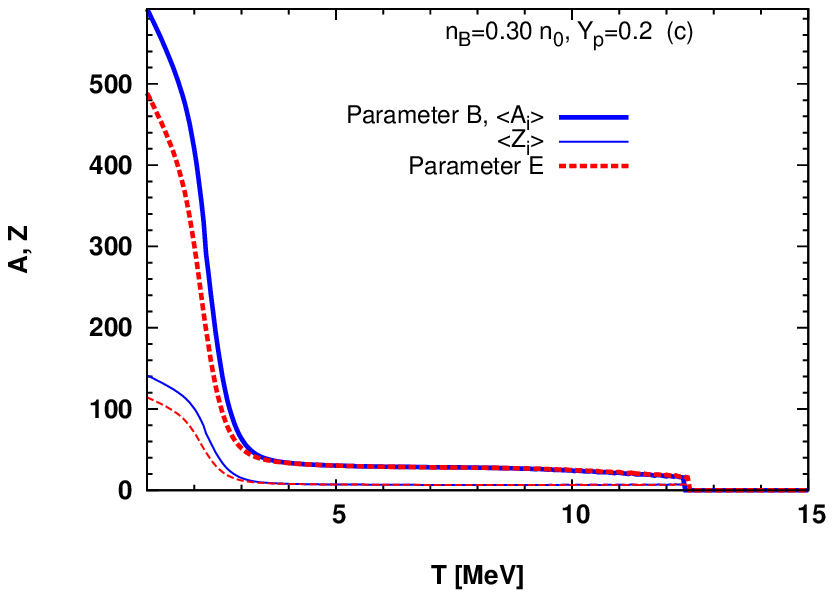}
\includegraphics[width=8.1cm]{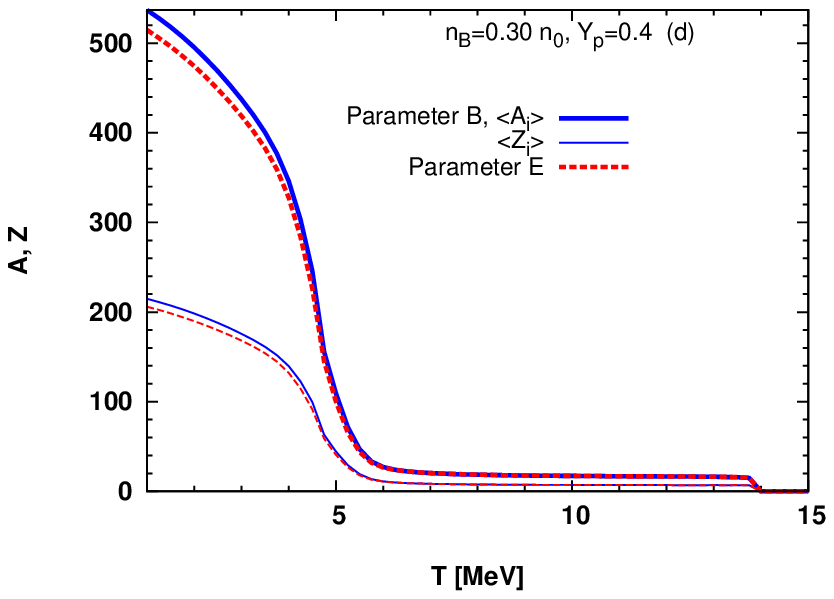}
\caption{Average mass  and proton numbers   (thick and thin lines) of heavy nuclei as  functions of  $T$ in the self-consistent model 
 for parameter sets  B (blue solid lines) and E (red dashed  lines)
at  $n_B=0.1 \ n_0$ (top row) and  $0.3  \ n_0$ (bottom row)
and $Y_p=$ 0.2 (left column) and 0.4 (right column).
}
\label{fig_mamz}
\end{figure}

\begin{figure}
\includegraphics[width=8.1cm]{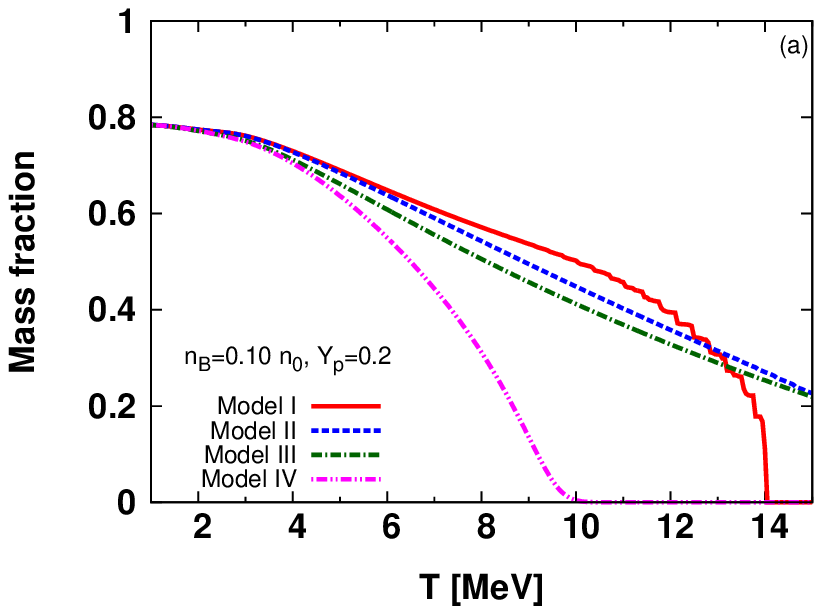}
\includegraphics[width=8.1cm]{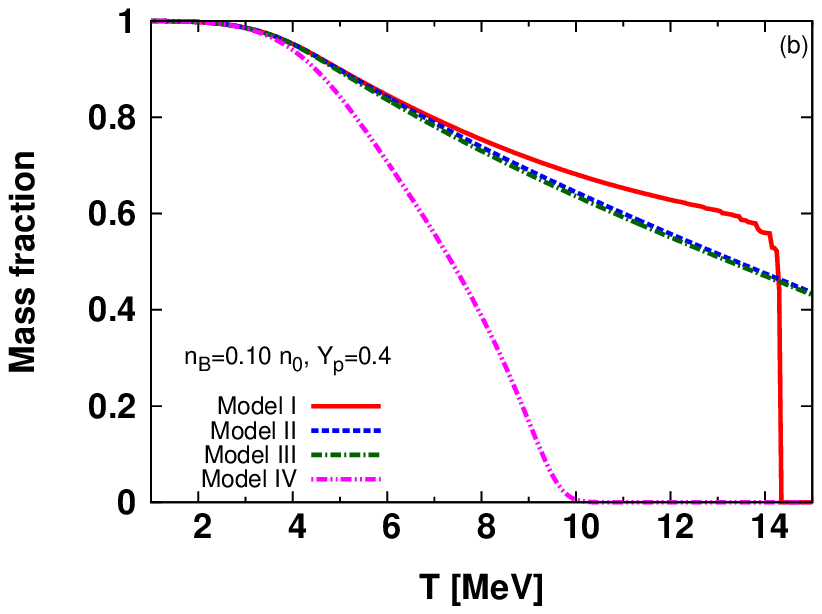}
\includegraphics[width=8.1cm]{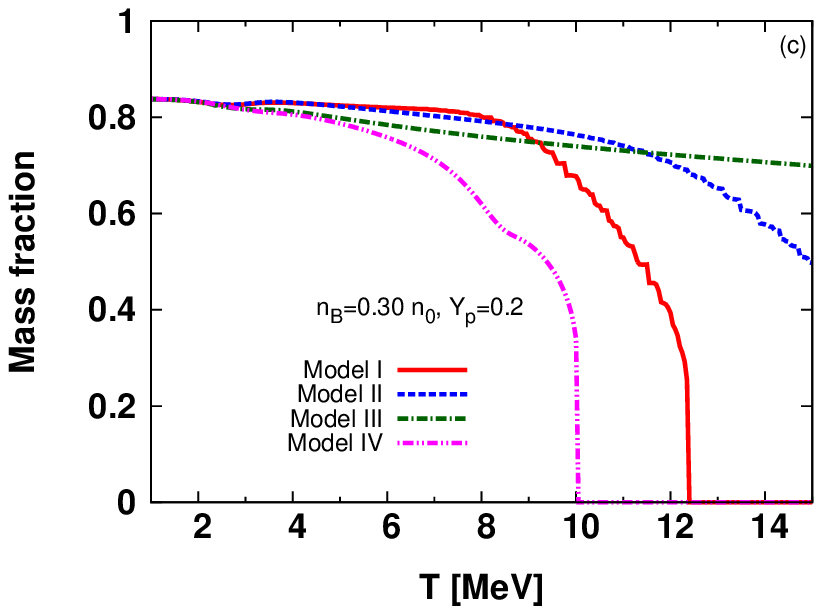}
\includegraphics[width=8.1cm]{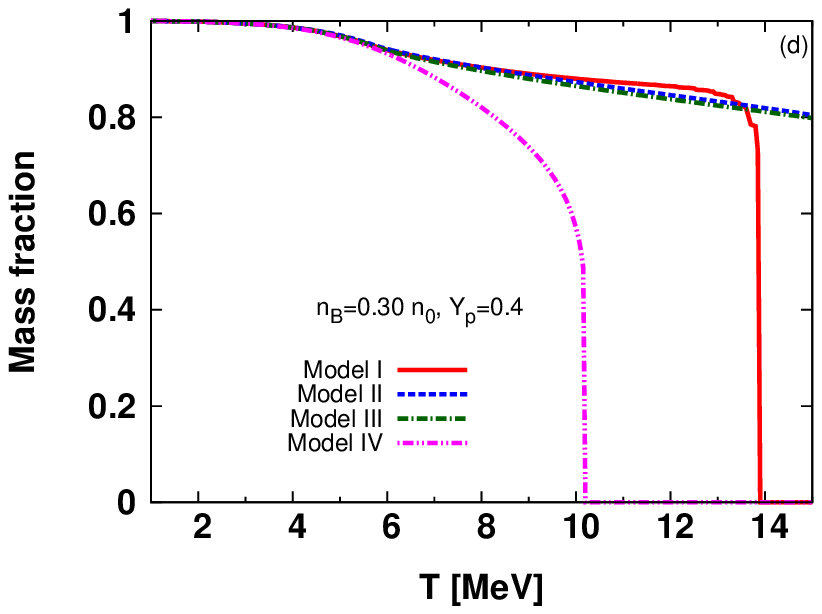}
\caption{Comparison of mass fractions of heavy nuclei  among different nuclear models listed on Table~\ref{tab2_model}, the self-consistent calculation with the compressible-liquid drop model (Model I, red solid lines),  the incompressible-liquid drop models  with the same $T_c(Z,N)$ of surface tensions  as in Model I (Model II, blue dashed lines) and with $T_c=18$ MeV for the all nuclei (Model III, green dashed-dotted lines) and the incompressible one with $T_c=\infty$ and the nuclear dissolution factor \cite{pais16}, Eq.~(\ref{eq:fla}), (Model IV, magenta dashed double-dotted  lines)
at  $n_B=0.1 \ n_0$ (top row) and  $0.3  \ n_0$ (bottom row)
and $Y_p=$ 0.2 (left column) and 0.4 (right column).
All the four models are calculated with the parameter set  B.
}
\label{fig_xfracmp}
\end{figure}


\begin{figure}
\includegraphics[width=8.1cm]{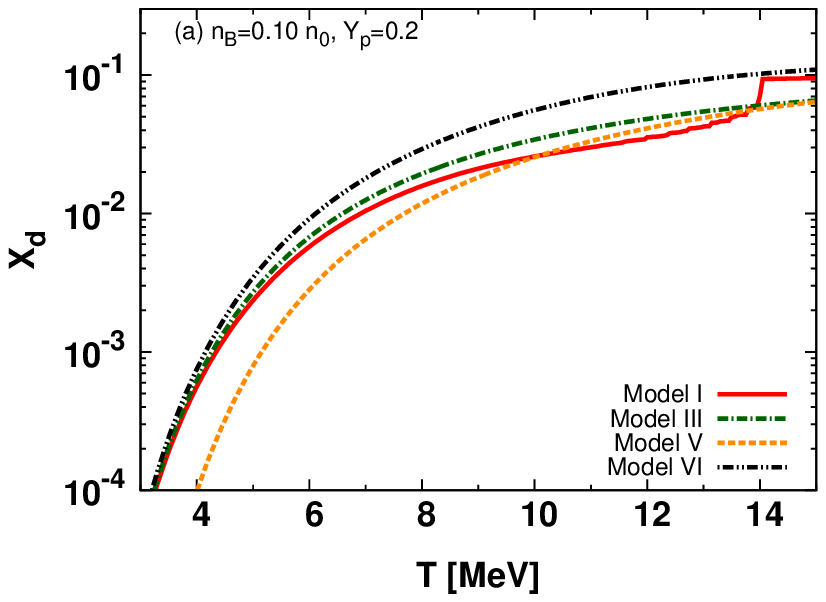}
\includegraphics[width=8.1cm]{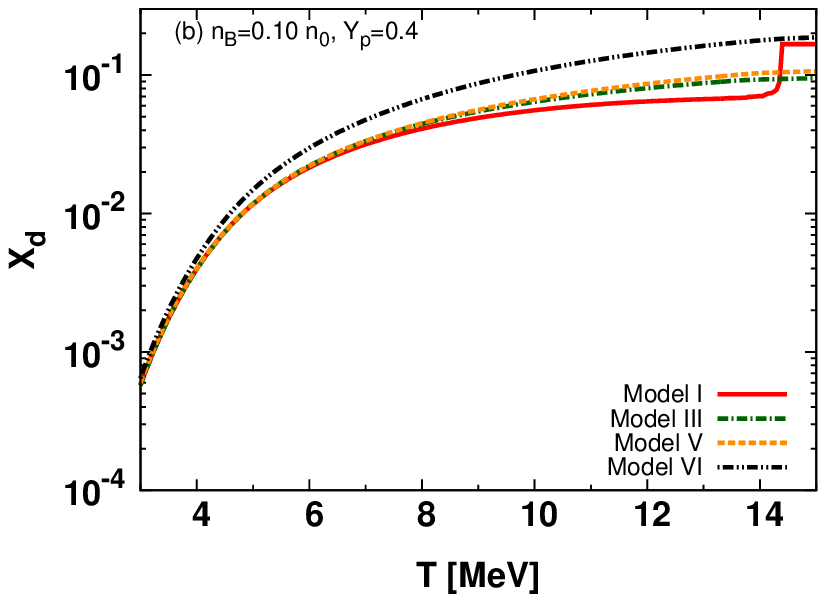}
\includegraphics[width=8.1cm]{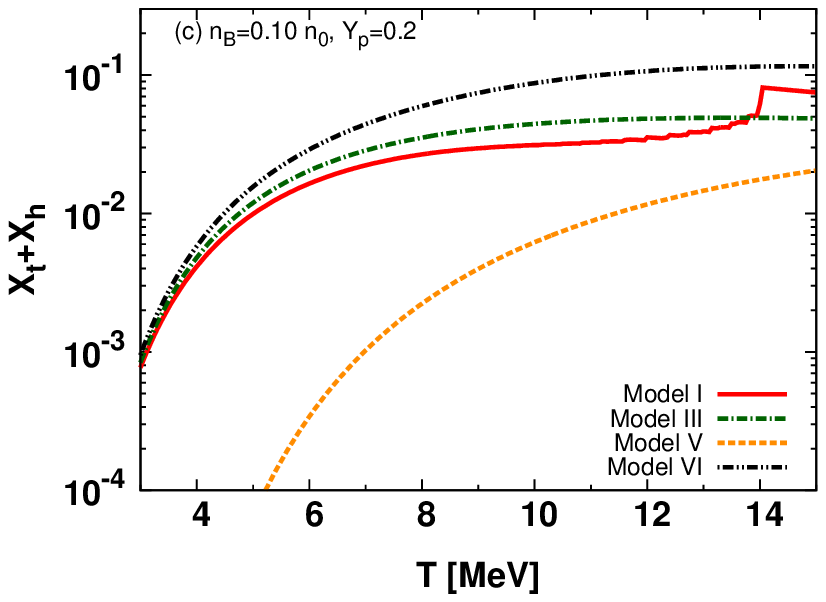}
\includegraphics[width=8.1cm]{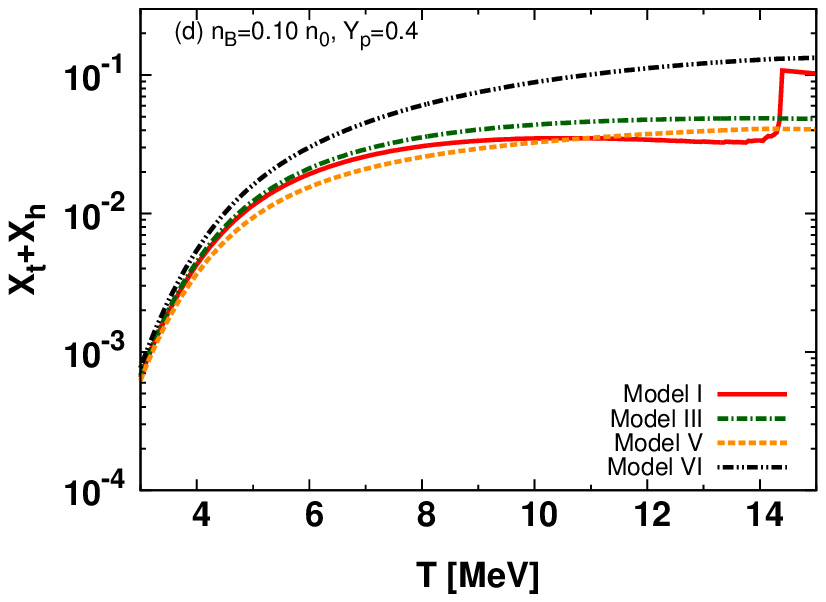}
\includegraphics[width=8.1cm]{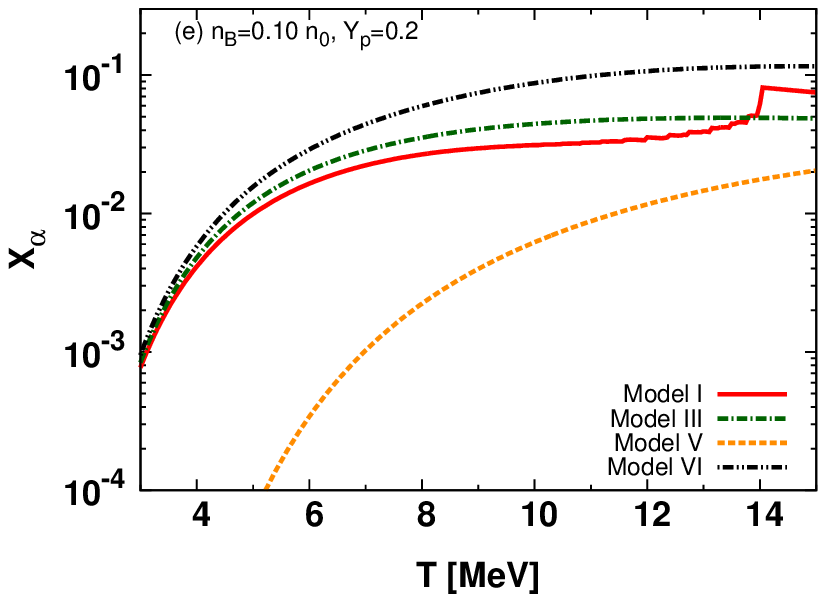}
\includegraphics[width=8.1cm]{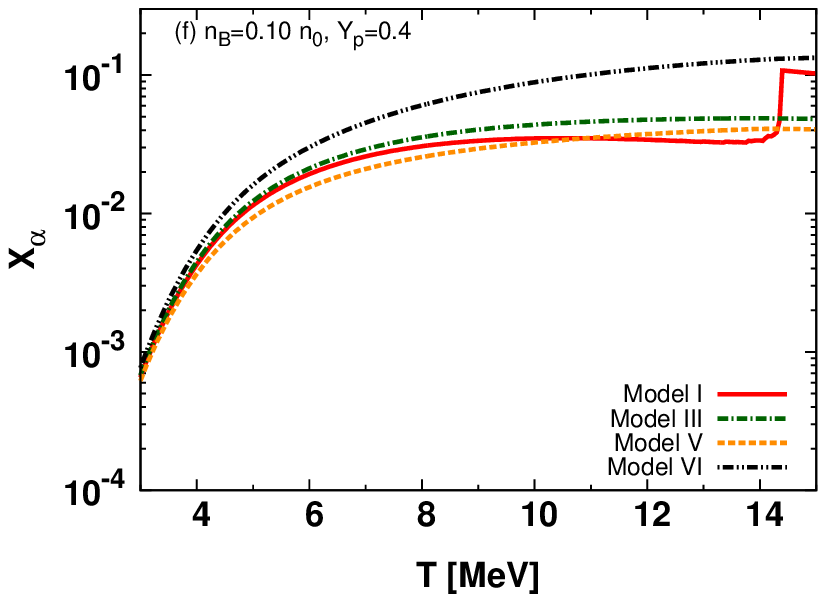}
\caption{Mass fractions of  deuteron (top row), triton and helion (middle row)  and alpha particle (bottom row)
for Model I (red solid lines),   Model III (green dashed-dotted lines), and  Model V (orange dashed lines) with a quantum approach to light clusters and Model VI (black dashed double-dotted  lines) with a level-density approach to light clusters
at  $n_B=0.1 \ n_0$ and $Y_p=$ 0.2 (left column) and 0.4 (right column).
The three models other than Model I employ the same incompressible liquid drop models for the heavy nuclei.
}
\label{fig_light}
\end{figure}

\end{document}